\documentclass[useAMS,usenatbib,usegraphicx,referee]{mn2e}
\title[Amount of Intergalactic Dust]
{Amount of intergalactic dust: constraints from distant
supernovae and thermal history of intergalactic medium}
\author[A. K. Inoue \& H. Kamaya]
{Akio K. Inoue\thanks{E-mail:akinoue@scphys.kyoto-u.ac.jp}\thanks{
JSPS Research Fellow}$^1$  and Hideyuki Kamaya$^2$\\ 
$^1$ Department of Physics, Faculty of Science, Kyoto University,
Sakyo-ku, Kyoto 606-8502, Japan\\
$^2$ Department of Astronomy, Faculty of Science, Kyoto University,
Sakyo-ku, Kyoto 606-8502, Japan}
\begin{document}

\date{}

\pagerange{\pageref{firstpage}--\pageref{lastpage}} \pubyear{2003}

\maketitle

\label{firstpage}

\begin{abstract}

This paper examines the allowed amount of IG (intergalactic) dust, which
is constrained by extinction and reddening of distant SNe Ia and thermal
history of IGM (intergalactic medium) affected by dust photoelectric
heating. Based on the observational cosmic star formation history, we
find an upper bound of $\chi$, the mass ratio of the IG dust to the
total metal in the Universe, as $\chi\la 0.1$ for $10 {\rm \AA} \la a
\la 0.1 \micron$ and $\chi\la 0.1(a/0.1\,\micron)$ for $0.1
\micron\la a\la1\micron$, where $a$ is a characteristic grain size of
the IG dust.  This upper bound of $\chi\sim0.1$ suggests that the
dust-to-metal ratio in the IGM is smaller than the current Galactic
value.  The corresponding allowed density of the IG dust increases from
$\sim10^{-34}$ g cm$^{-3}$ at $z=0$ to $\sim10^{-33}$ g cm$^{-3}$ at
$z\sim1$, and keeps almost the value toward higher redshift.
This causes IG extinction of $\la 0.2$ mag at the observer's $B$-band
for $z\sim 1$ sources and that of $\la 1$ mag for higher redshift
sources.  Furthermore, if $E(B-V)\sim 0.1$ mag at the observer's frame
against $z\ga1$ sources is detected, we can conclude that a typical size
of the IG dust is $\la 100$ \AA.  The signature of the 2175 \AA\ 
feature of small graphite may be found as a local minimum at 
$z\sim2.5$ in a plot of the observed $E(B-V)$ as a function of the 
source redshift. Finally, the IGM mean temperature at $z\la1$ can be 
still higher than $10^{4}$ K, provided the size of the IG dust is 
$\la100$ \AA.

\end{abstract}


\section{Introduction}

As long as there is dust between radiation sources and observers, the
dust extinction and reddening\footnote{In this paper, we call the total
absolute amount of the absorption and scattering at a wavelength just
$extinction$, and the differential extinction between two wavelengths
$reddening$.} must be corrected if we want to realize the nature of the
sources.  We should examine how much extinction and reddening there are.
The extinction property in the Galaxy is a well studied example (e.g.,
\citealt{dra84}). 
Using HI and far-infrared emission as tracers for the
dust column density, we can obtain the extinction amount by the
Galactic dust with reasonable accuracy \citep{bur82,sch98}. 
Although the dust distribution and
properties in the external galaxies are not well known yet, we can
correct the dust extinction in the galaxies by using some indicators of
the extinction with some assumptions (e.g., \citealt{bua02}).  How about
the extinction by the intergalactic (IG) dust?

We have already known the fact that some metal elements exist in the
Lyman $\alpha$ clouds even at redshift larger than 3 (e.g.,
\citealt{cow95,telf02}). It suggests that the dust grains also exist in
the low-density intergalactic medium (IGM). Such diffuse IG dust causes
the IG extinction and reddening, which may affect on our understanding
of the Universe significantly. One might think that the IG dust amount
is negligible because such a significant IG reddening is not reported in
the previous studies \citep{tak72,che91,rie98,per99}.  However, the
wavelength dependence of the IG extinction is quite uncertain. If it is
gray as suggested by \cite{agu99}, a large extinction is possible with no
reddening. Nobody can conclude that the IG dust is negligible because of
no observable reddening.

Theoretically, it is predicted that metals synthesized in supernova
(SN) explosions form into the dust grains in the cooling ejecta of SNe
\citep{koz87,tod01,noz03,sch03}. Recently, 
thermal emissions of such dust from two supernova remnants, Cas A and
Kepler, are detected \citep{dun03,morg03}. In a very high-$z$ universe,
SNe of massive Population III stars formed in low mass halos, which are
likely to be the main site of the first star formation, can disperse the
produced metals into the IGM \citep{bro03}. The dust grains may be also
dispersed into the IGM. Therefore, the IG dust grains may exist even in
a $z\ga10$ universe \citep{elf03}.   

Extinction by the IG dust may affect on the determination of the
cosmological parameters from observations of SNe. The observed
dimming beyond the geometrical dimming in the empty space of distant
($z\simeq 0.5$) Type Ia SNe, which is attributed to the cosmological
constant \citep{rie98,per99}, can be reproduced by the gray IG
extinction without the cosmological constant \citep{agu99}. \cite{goo02}
show that the apparent brightening of the farthest SNe Ia ($z=1.7$;
\citealt{rie01}) can be also explained by the gray IG extinction with
zero cosmological constant if the dust-to-gas ratio in the IGM decreases
properly with increasing redshift. Therefore, to know the amount of the
IG dust is also important in the cosmological context.

The evidence of the IG dust should be imprinted in the cosmic microwave
background (CMB) and infrared background because the dust emits thermal
radiation in the wave-band from the far-infrared to submillimetre (submm)
\citep{row79,wri81,elf03}. Although the {\it COBE} data provides us with
only a rough upper bound on the IG dust \citep{loe97,fer99,agu00}, the data
of {\it WMAP} \citep{spe03} may be promising. The submm background
radiation will give a more strict constraint on the IG dust 
\citep{agu00}.

Recently, we have proposed a new constraint on the IG dust amount by
using thermal history of the IGM \citep{ino03}.  Since the dust
photoelectric heating is very efficient in the IGM \citep{nat99}, the
theoretical thermal evolution of the IGM taking into account of the
heating by dust violates the observational temperature evolution if too
much IG dust is input in the model.  Hence, \cite{ino03} obtain an upper
bound of the IG dust amount in order that the theoretical IGM
temperature should be consistent with the observed one. The obtained
upper bound of the dust-to-gas ratio in the IGM is 1\% and 0.1\% of the
Galactic one depending on the IG grain size of $\sim 100$ \AA -- 0.1
$\mu$m and $\sim 10$ \AA, respectively, at redshift of $\sim 3$. 

In this paper, with help of distant SNe Ia observation, we extend our
previous approach in order to discuss the upper bounds of the IG dust
extinction and reddening.  In the next section, we start from the cosmic
star formation history (SFH) to specify the IG dust amount at each
redshift.  According to the assumed SFH, we can estimate IG dust
extinction and reddening at each redshift theoretically.  In section 3,
we comment on observational constraints from the extinction and
reddening of distant SNe Ia.  In section 4, further constraints are
presented by comparing theoretical and observational thermal histories
of the IGM.  Based on the allowed amount of the IG dust, we also discuss
some implications from our results in section 5.  The achieved
conclusions are summarised in the final section.

Throughout the paper, we stand on a $\Lambda$-cosmology.  That is, we
constrain the amount of the IG dust in order that the IG dust should not
affect on the determination of the cosmological parameters from distant
SNe Ia. This is because the flat universe is favored by results of CMB
observations \citep{jaf01,pry02,spe03}, whereas only the matter cannot
make the flat universe \citep{per01}. Furthermore, the recent
observations of the X-ray scattering halo around high-$z$ QSOs suggest
too small amount of the IG dust to explain all amount of the dimming of
the distant SNe Ia \citep[but see also \citealt{win02}]{pae02,tel02}.
\cite{mor03} also reach the same
conclusion by analyzing the observed colours of the SDSS (Sloan Digital
Sky Survey) quasars.  The following cosmological parameters are adopted:
$H_0=70$ km s$^{-1}$ Mpc$^{-1}$, $\Omega_{\rm M}=0.3$,
$\Omega_\Lambda=0.7$, and $\Omega_{\rm b}=0.04$.

\section{Star formation history and intergalactic dust}

To estimate the IG extinction and reddening theoretically, we must
investigate production of dust at each redshift.  Since dust is made of
metals, a cosmological evolution of the metal amount should be
specified. As metals are products of stellar evolution,
therefore, we shall specify the cosmic SFH as a first step.

Since \cite{mad96}, researches of the cosmic SFH are extensively
performed. In figure 1, we show observational star formation rates in a
unit comoving density as a function of redshift.  The cross and open
symbols are estimated from the H$\alpha$ line and the rest-frame
ultra-violet (UV) luminosities not corrected by the interstellar dust
extinction. Hence these are lower limits. The filled-circles are
estimated from the submm data. Due to the small statistics, the
uncertainty of the submm data is rather large.  
The real SFH is still uncertain because we do not know the suitable
correction factor against the internal dust extinction.
In this paper, therefore, we adopt two example models: high and low
SFHs, which are shown in figure 1 as solid and dashed curves,
respectively. For the low SFH case, we have employed a conservative
correction factor for the internal dust extinction. The high SFH case is
set to be compatible with the submm data. Quantitatively, these SFHs are
formulated as 
\begin{equation}
 \frac{\rho_{\rm SFR}^*(z)}{0.1 M_\odot\,{\rm yr}^{-1}\,{\rm Mpc}^{-3}} = 
  \cases{
   \left(\frac{1+z}{2}\right)^{3.3} & (for $z\leq 1$) \cr
   \left(\frac{1+z}{2}\right)^{-1.5} & (for $z>1$ low SFH) \cr
   1 & (for $z>1$ high SFH) \cr
   }\,,
\end{equation}
where $\rho_{\rm SFR}^*$ is the star formation rate per unit comoving
density. Recently, some authors suggest that a SFH like the high case is
more likely \citep[e.g.,][]{spr03,gia04}. 
Hence, we discuss the high SFH case mainly in the following.

\begin{figure}
 \includegraphics[height=8.0cm,clip] {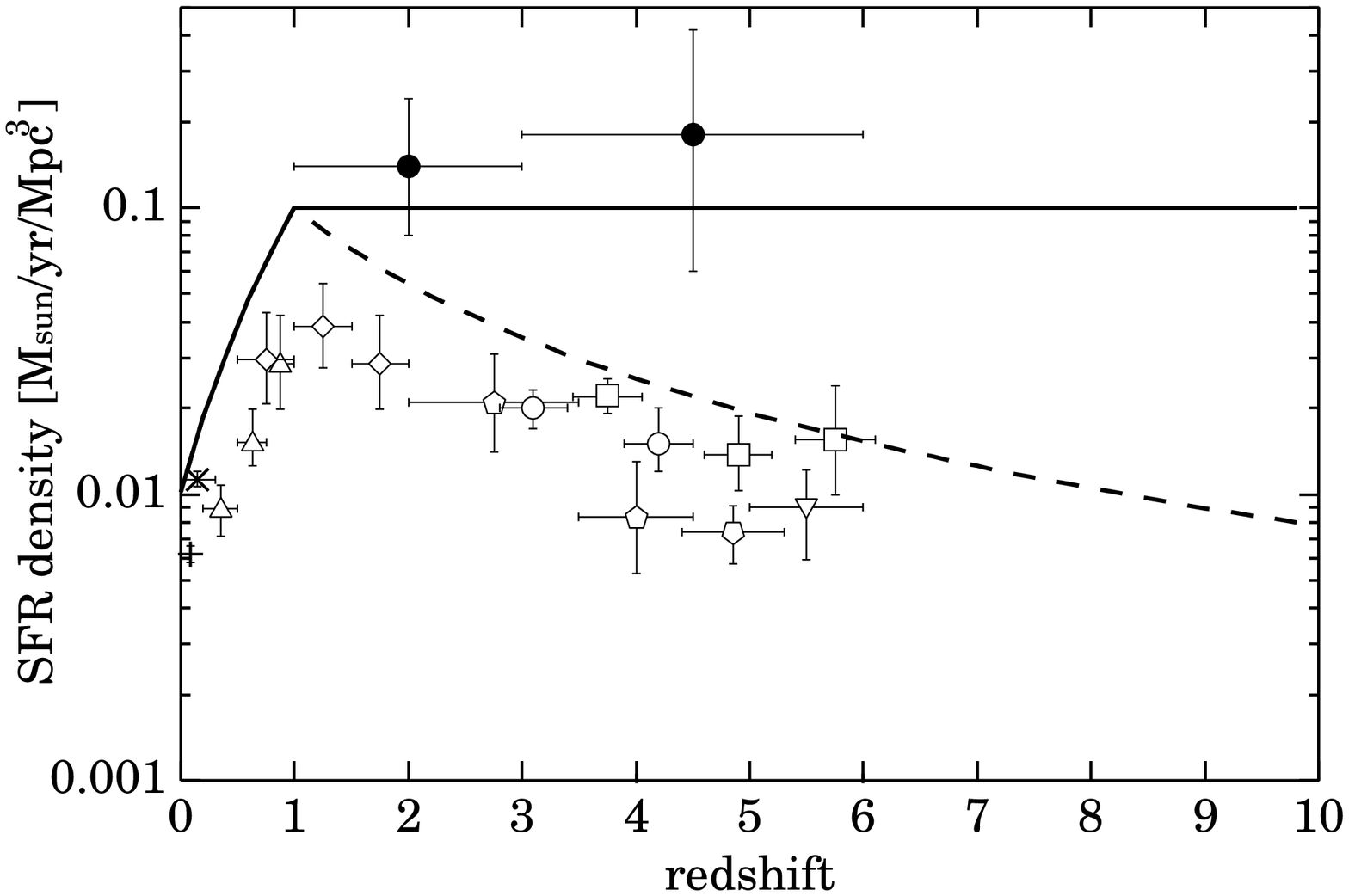}
 \caption{Cosmic star formation rate density as a function of
 redshift. The plotted points with error-bars are estimated from the
 H$\alpha$ data: cross \citep{gal95} and inclined cross \citep{tre98}; 
 from the rest-frame ultra-violet data: open-triangles \citep{lil96},
 open-diamonds \citep{con97}, open-pentagons \citep{mad98}, open-circles
 \citep{ste99}, open-squares \citep{gia04}, open-inverse-pentagon
 \citep{iwa03}, open-inverse-triangle \citep{bou03}; and from the
 submillimetre data: filled-circles \citep{bar00}. These points are
 adjusted to the cosmological parameters assumed in this paper. The
 solid and dashed curves are the models of high and low star formation
 histories, respectively.}
\end{figure}

Once a SFH is specified and if the instantaneous recycling approximation
is adopted, the cosmic mean metallicity\footnote{In this paper, we
define metallicity as the mass fraction of elements heavier than Li as
the usual way.} evolution is determined by:
\begin{equation}
 Z(z) = \frac{y_Z}{\Omega_{\rm b}\rho_{\rm c,0}} 
  \int_z^{z_{\rm S}} \rho_{\rm SFR}^*(z') \frac{dz'}{H(z')(1+z')}\,,
\end{equation}
where $y_Z$ is the produced metal mass fraction for a unit star forming
mass, $\rho_{\rm c,0}$ is the current critical density, $z_{\rm S}$ is
the starting redshift of the cosmic star formation, and $H(z)$ is the
Hubble constant at the redshift $z$. We have assumed $y_Z$ to be a
constant for simplicity. If the Salpeter initial mass function (0.1--125
$M_\odot$) is assumed, $y_Z=0.024$ \citep{mad96}.  We assume $z_{\rm
S}=10$ in this paper. This does not affect on the results obtained in
the following sections because the measure of time along the
redshift is small in the high-$z$ universe.  Indeed, the
cosmic metallicities in $z\la 3$ for various $z_{\rm S}$ become nearly
equal to each other if $z_{\rm S}\ga 5$.  

In figure 2, we show the mean cosmic metallicity as a function of
redshift.  In the figure, the lower limit of metallicity in the
Ly$\alpha$ clouds measured from C IV and Si IV by \cite{son01}
and the ranges of carbon abundance obtained by \cite{schaye03}
and oxygen abundance obtained by \cite{telf02} are overlaid. 
Our theoretical estimate of total metallicity in the universe is
compatible with the observed oxygen abundance in the IGM. This may mean
that a large part ($\sim50$\%) of metal produced in galaxies exist out
of galaxies. 
On the other hand, our estimate is much larger than the observed carbon
abundance. This may indicate that carbon is not suitable to trace the
cosmic mean metallicity in the early phase since the first metal
pollution is made by the Type II SNe  \citep[e.g.,][]{pag97}.  
Indeed, \cite{agu03} find [Si/C]$\sim 0.8$. 

\begin{figure}
 \includegraphics[height=8.0cm,clip] {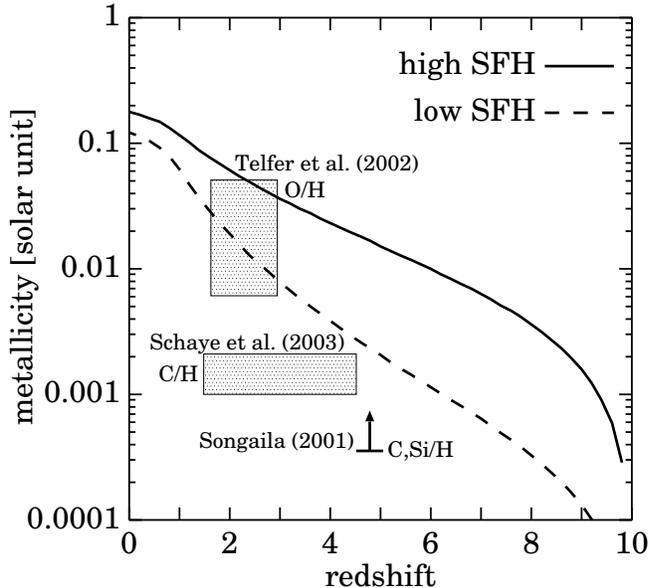}
 \caption{Mean cosmic metallicity evolution. The vertical axis is
 normalized by the Solar metallicity $Z_\odot=0.02$. The solid and
 dashed curves are the cases of the high and low star formation
 histories, respectively. The starting redshift of the cosmic star
 formation is set to be $z_{\rm S}=10$. The observed range of oxygen
 abundance by \citet{telf02}, carbon abundance by \citet{schaye03}, and
 lower limit of metallicity obtained by \citet{son01} are overlaid as
 the shaded squares and the upward arrow, respectively. 
 We note that \citet{agu03} find [Si/C]$\sim 0.8$.}
\end{figure}

Let us introduce one parameter to describe the amount of the IG dust;
the ratio of the IG dust mass to the total metal mass in the universe
defined as 
\begin{equation}
 \chi \equiv \frac{\rm IG\, dust\, mass}{\rm total\, metal\, mass} 
  = \frac{{\cal D}^{\rm IGM}}{Z}\,,
\end{equation}
where ${\cal D}^{\rm IGM}$ is the dust-to-gas mass ratio in the IGM.  In
principle, this parameter is determined by the transfer mechanism of
dust grains from galaxies into the IGM.  Although some authors have
tried to solve this problem approximately 
(e.g., \citealt{fer90,fer91,agu01}), the results
are not still conclusive.  This is because we must solve problems of the
magneto-radiation-hydrodynamics of dusty plasma finally.  Here we
approach the parameter $\chi$ by another way; we constrain the parameter
by using the observational data of distant SNe Ia (\S3) and the thermal
history of the IGM (\S4). While the parameter $\chi$ may evolve along
redshift, we treat it as a constant for simplicity. Hence, the obtained
values of $\chi$ in the following sections are regarded as those
averaged over redshift.

A dust model must also be specified.  In this paper, we adopt the
``graphite'' and the ``smoothed astronomical silicate'' models by
\cite{dra84,lao93,wei01a}, which can explain the interstellar dust
properties in the Galaxy and the Magellanic clouds very well. Although
there is no evidence that the IG dust is the same as the Galactic dust,
we assume them as a working hypothesis. 

The grain size distribution in the IGM is also quite uncertain. 
As a first step, we assume all grains to have the same size. 
This means that the grain size in the current paper indicates a
characteristic size of the IG dust (i.e. an averaged size by a certain
way). 
\cite{agu99} suggests a selection rule in the transfer of
dust grains from the host galaxies into the IGM; the small grains are
destroyed by the thermal sputtering process when the grains are
transfered through the hot gas halo of the host galaxies. Indeed, this
theoretical suggestion is very interesting to realize the gray dust
model.  However, we also examine the possibility of the small IG grain
because the suggestion of the selection rule is not confirmed
observationally at the moment.

\section{Constraint from observational dimming of supernovae}

In this section, we constrain the amount of the IG dust by means of
observational dimming of distant Type Ia SNe.  
According to \cite{rie01}, the dimming of SNe Ia 
at $z\sim 0.5$ is $\sim 0.2 \pm 0.1$ mag against the
empty universe.  Our policy is that the IG dust extinction does not
affect the interpretation of the cosmological constant.  Thus, we
attribute the 0.2 mag dimming of SNe Ia at $z\sim 0.5$ to the cosmological
constant and regard the quoted error (0.1 mag) as an upper limit of the 
contribution by the IG dust to the dimming.
Although the cosmological dimming does not depend on the observed
wavelength, the dimming by the IG dust may depend on the wavelength.
As the distant SNe Ia are observed in $B$ and $V$-bands and  $B$-band
provides a slightly more strict constraint of the IG dust than $V$-band,
we regard the upper limit of 0.1 mag as that in the $B$-band. 
In addition, \cite{per99} report that the difference of the
observed reddening between the local and distant SNe is $\langle
E(B-V)\rangle_{z=0.5}-\langle E(B-V)\rangle_{z=0.05} =0.002\pm0.03$
mag. Although there seems to be no systematic difference, we can still
consider the absolute value of the colour excess by the IG dust less
than 0.03 mag.  In summary, the IG extinction and
reddening from observations of distant SNe Ia at $z\sim 0.5$ are $A^{\rm
IGM}_{B,z=0.5}\leq0.1$ mag and $|E(B-V)^{\rm IGM}_{z=0.5}|\leq0.03$ mag. 

Suppose an observer who observes a source with a redshift of $z$ at a
wavelength of $\lambda_{\rm obs}$ in his/her rest frame.  The amount of
the IG extinction is given by
\begin{equation}
 \frac{A_{\lambda_{\rm obs}}^{\rm IGM}(z)}{\rm mag} = 
  1.086 \pi a^2 \int_0^z Q\left(a,\frac{\lambda_{\rm obs}}{1+z'}\right) 
  n_{\rm d}^{\rm IGM}(z') \frac{c\, dz'}{(1+z')H(z')}\,,
\end{equation}
where $a$ is the grain radius, $Q$ is the extinction efficiency factor,
$c$ is the light speed, and $n_{\rm d}^{\rm IGM}$ is the IG grain number
density in a unit physical volume, which is
\begin{equation}
 n_{\rm d}^{\rm IGM}(z)=\frac{\chi \Omega_{\rm b}\rho_{\rm c,0}(1+z)^3 Z(z)}
  {4\pi a^3 \varrho/3}\,,
\end{equation}
where $\varrho(=2\,{\rm g\,cm^{-3}})$ is the grain material density.
For simplicity, we have assumed that dust grains distribute uniformly
with $n_{\rm d}^{\rm IGM}$ corresponding to the redshift. We do not
consider any structure of the dust distribution.

In figure 3, we show the amount of the IG extinction divided by $\chi$
for a source with $z=0.5$ as a function of the assumed grain size. While
the results for the high SFH is depicted in the figure, the extinction
amount for the low SFH case is only a factor of about 1.5 smaller than
that of the high SFH case. 
The extinction amount is independent of the grain size as long as 
$2\pi a < \lambda$. This is caused by  
(1) the extinction cross section ($=Q_{\rm ext}\pi a^2$) is proportional
to $a^3$ because the extinction efficiency factor, $Q_{\rm ext}$, is
proportional to the grain size, $a$, linearly, and (2) the number
density of the grain has a dependence of $a^{-3}$ for a fixed dust mass.
On the other hand, $Q_{\rm ext}$ becomes constant (almost 2) when 
$2\pi a > \lambda$, so that the extinction cross section is determined
by mainly the geometrical one which is proportional to $a^2$.  Thus, 
the extinction amount is proportional to $a^{-1}$.  When
$2\pi a \sim \lambda$, grains interact with photons the most
effectively, so that the amount of extinction shows a peak in both
panels of figure 3. 

\begin{figure}
 \includegraphics[height=8.0cm,clip] {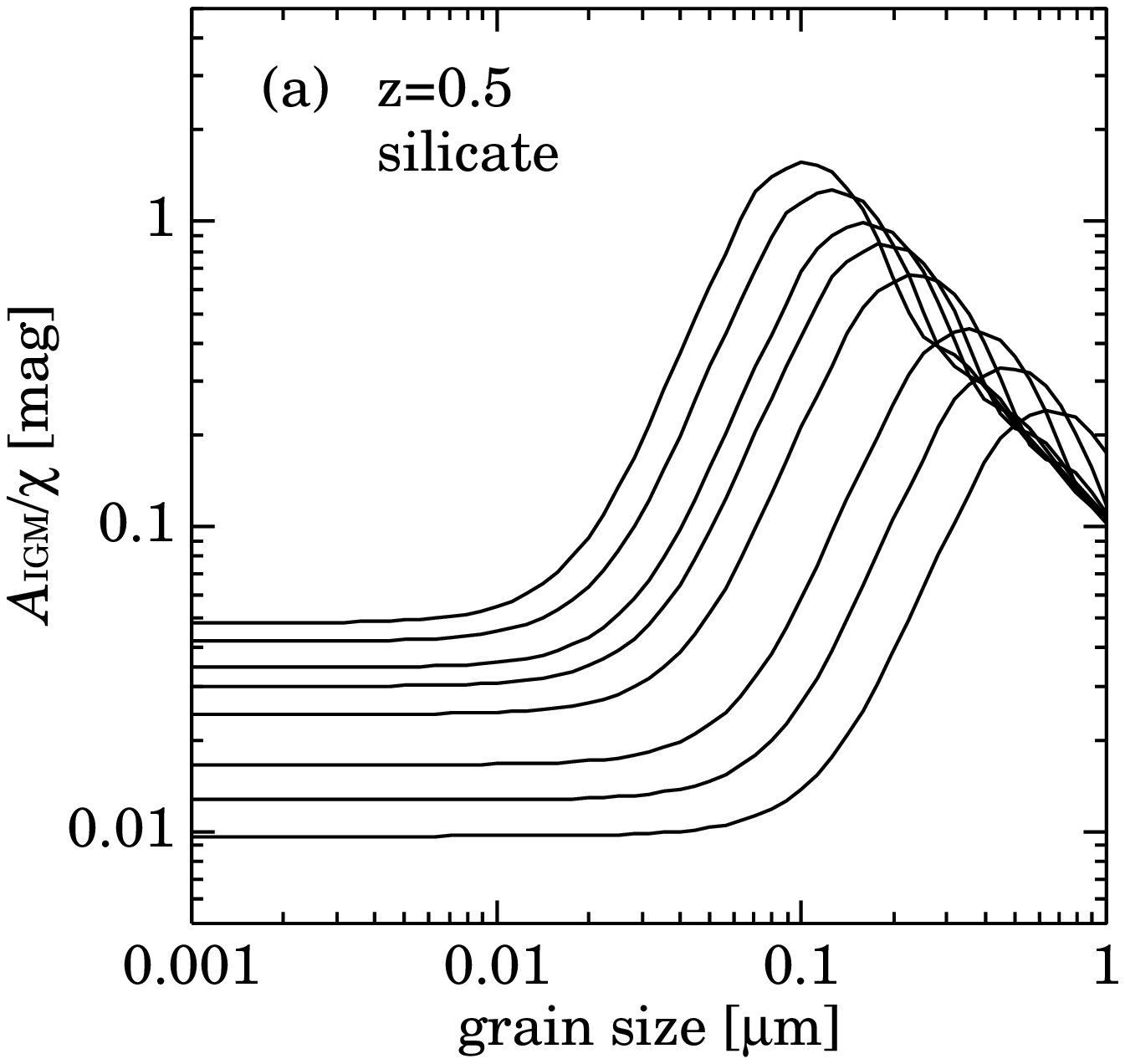}
 \includegraphics[height=8.0cm,clip] {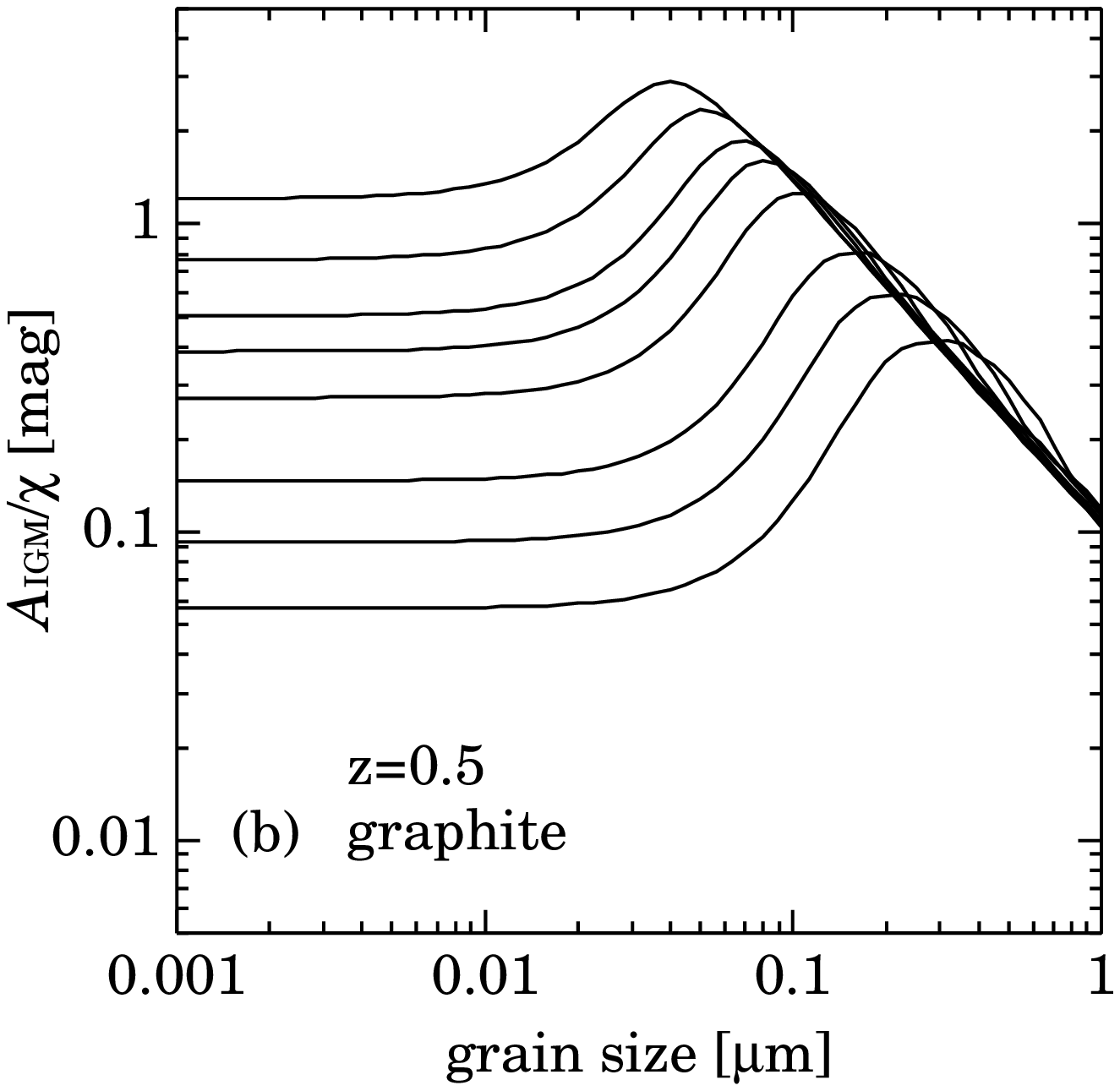}
 \caption{Intergalactic extinction in various bands against a source
 with the redshift $z=0.5$ as a function of the IG grain size. The
 vertical axis is divided by the parameter $\chi$, which is defined
 by equation (3). The panels (a) and (b) are the cases of silicate and
 graphite, respectively. The star formation history assumed is the lower
 case shown in figure 1. The solid curves indicate the extinction amount
 in the $U$, $B$, $V$, $R$, $I$, $J$, $H$, and $K$ bands from top to
 bottom.}
\end{figure}

We shall remember the observational constraints of $A_{B,z=0.5}^{\rm
IGM}\leq 0.1$ mag and $|E(B-V)^{\rm IGM}_{z=0.5}| \leq 0.03$ mag.
Hence, we can obtain the upper bound of $\chi$ via Eq.(4) or figure 3,
which is shown in figure 4.  The silicate and graphite cases are shown
in the panel (a) and (b), respectively. 
The dotted and dashed curves indicate the
upper bounds of $\chi$ based on $A_{B,z=0.5}^{\rm IGM} = 0.1$ mag and
$|E(B-V)^{\rm IGM}_{z=0.5}| = 0.03$ mag, respectively. 
The discontinuity of the dashed curve is due to $E(B-V)=0$; the left of
the discontinuity is positive $E(B-V)$ and the right is negative one.
The solid curve is the upper bound based
on the thermal history of the IGM which is obtained in the next
section. The rejected area of $\chi$ is shaded; the thin shade means the
rejected area based on the SNe Ia extinction/reddening, and the thick shade
is the area based on the IGM thermal history (section 4). 
We show only the high SFH case. 

We find that, for both grain types, the upper bound of $\chi$ from the
distant SNe Ia is $ \sim \ 0.1(a/0.1\,{\rm \mu m})$ for 
$0.1\,{\rm \mu m} \la a \la 1\,{\rm \mu m}$.
While we have no constraint of $\chi$ for the small ($a \la 100$ \AA)
silicate grain (panel [a]), the upper bound of $\chi$ is $\sim 0.1$ when
$a \la 100$ \AA\ for the graphite case (panel [b]).  This difference is
caused by the different optical properties between graphite and
silicate; small silicate is more transparent than graphite in the
optical bands.  For $100\,{\rm \AA} \la a \la 0.1\,{\rm \mu m}$, the
upper bound of $\chi$ shows a local minimum corresponding to the peak
shown in figure 3. Finally, there is not any constraint of
$\chi$ for a very large ($a \ga 1$ $\mu$m) grain.

\begin{figure}
 \includegraphics[height=8.0cm,clip] {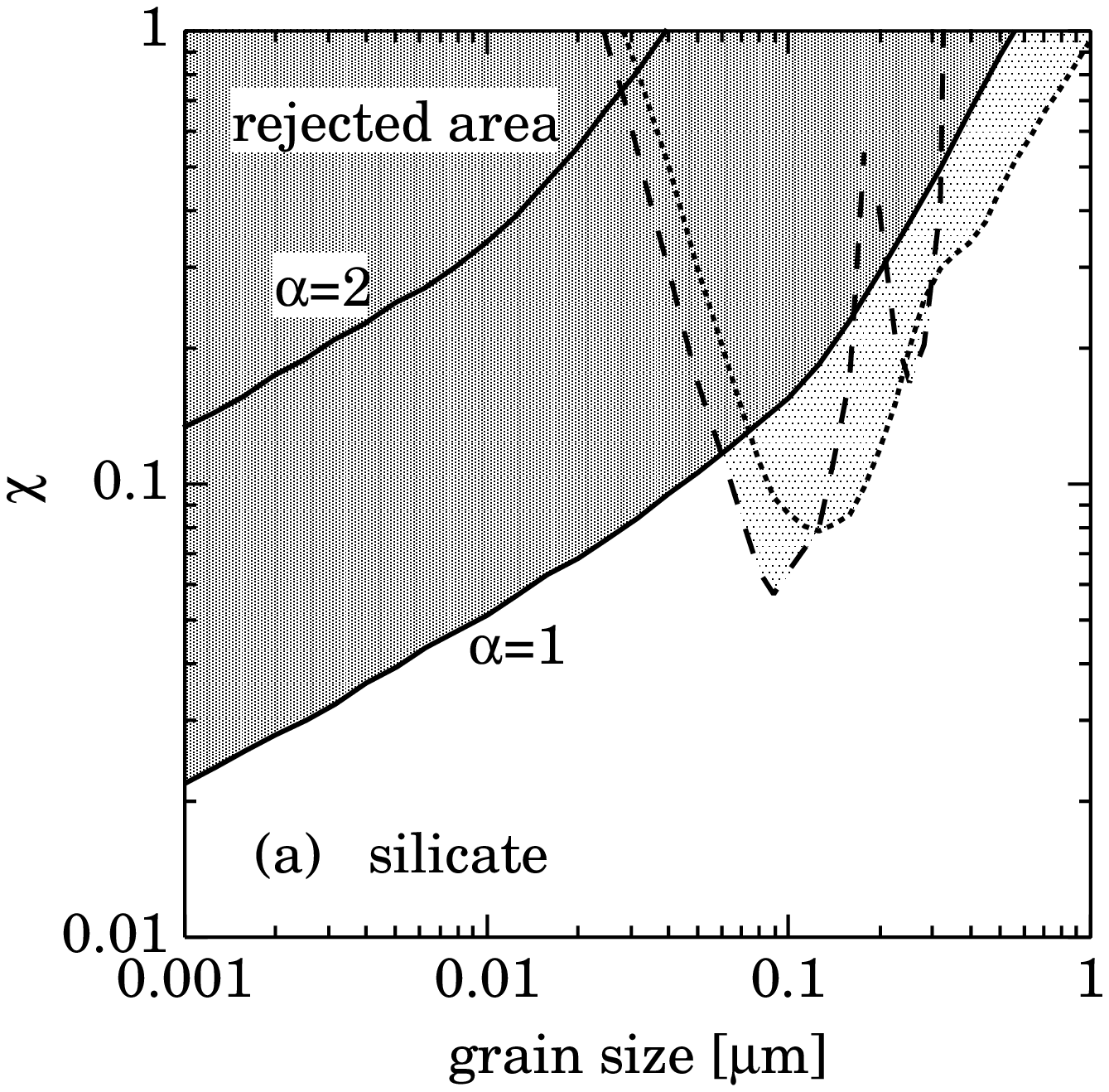}
 \includegraphics[height=8.0cm,clip] {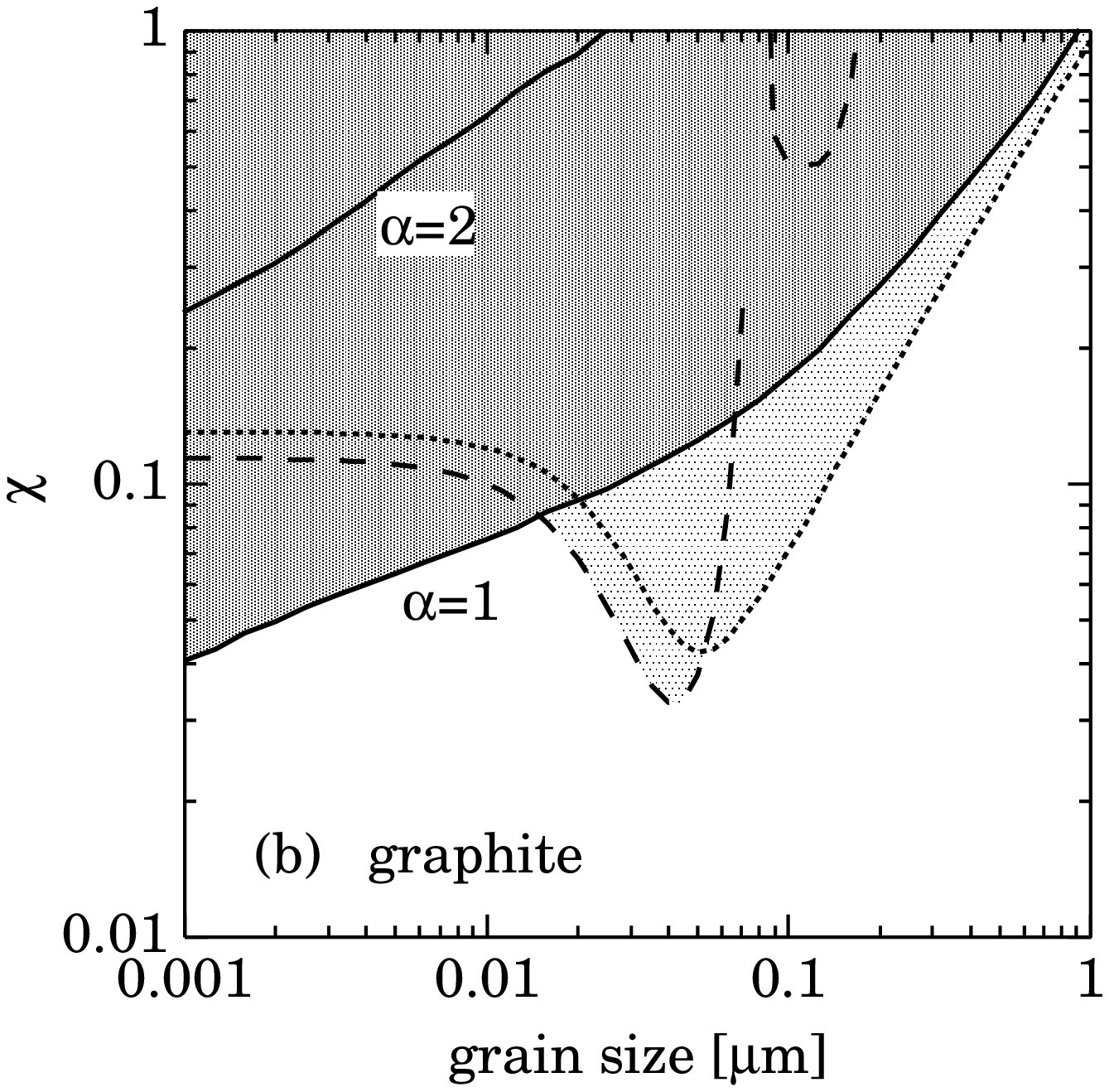}
 \caption{Rejected area of $\chi$, the mass ratio of the IG dust to
 the total metal as a function of the IG grain size: (a) silicate and
 (b) graphite case. The dotted and dashed curves are the upper bounds
 from extinction and reddening of distant SNe Ia. The solid curves are the
 upper bound obtained from the thermal history. Two spectral indices of 
 the background radiation are considered; $\alpha=1$ (bottom solid) and
 2 (top solid). The thick and thin shaded areas are the rejected areas
 based on the thermal history ($\alpha=1$ case) and on observations of
 SNe Ia, respectively. The high SFH is assumed.} 
\end{figure}

\section{Constraint from IGM thermal history}

As shown in the previous paper (\citealt{ino03}), we show that the
amount of the IG dust is constrained by using the thermal history of the
IGM. When a dust grain is hit by a photon with an energy larger than a
critical value, an electron escapes from the grain; photoelectric
effect. Such a photoelectron contributes to the gas heating if its
energy is larger than the mean kinetic energy of gas particles. As shown
in Appendix A (see also \citealt{nat99,ino03}), the photoelectric
heating by dust grains is comparable with, and sometimes dominate, the
atomic photoionization heating in the IGM. Of course, the efficiency of
the dust heating depends on the dust amount.  If a model of the IGM has
too much dust, the theoretical temperature of IGM exceeds the
observational one owing to the dust photoelectric heating.  Therefore,
we can put an upper bound of the amount of the IG dust so as to keep the
consistency between the theoretical IGM temperature and the observed
one. In the next subsection, we describe how to calculate the thermal
history of IGM affected by the dust photoelectric heating. An upper
bound of $\chi$, which represents the amount of the IG dust, is
estimated in subsection 4.2.

\subsection{Thermal history of IGM}

In this paper, we consider only a mean temperature of the IGM, $T_{\rm IGM}$, 
for simplicity. 
The $T_{\rm IGM}$ time-evolution is described by 
(e.g., \citealt{hui97})
\begin{equation}
 \frac{dT_{\rm IGM}}{dt}
  =-2HT_{\rm IGM} -\frac{T_{\rm IGM}}{X}\frac{dX}{dt}
  +\frac{2(\Gamma-\Lambda)}{3k_{\rm B}Xn_{\rm b}}\,,
\end{equation}
where $H$ is the Hubble constant, $n_{\rm b}$ is the cosmic mean number
density of baryon, $\Gamma$ and $\Lambda$ are the total heating and
cooling rates per unit volume, respectively, $X$ is the number ratio of
the total gaseous particles to the baryon particles, i.e., $X \equiv
\sum n_{i}/n_{\rm b}$, where $n_{i}$ is the number density of the $i$-th
gaseous species and we consider H~{\sc i}, H~{\sc ii}, He~{\sc i},
He~{\sc ii}, He~{\sc iii}, and electron. We neglect the effect of helium
and metal production by stars on the chemical abundance for simplicity.
Fortunately, time evolution of their abundance is not important. 
Indeed, the metal mass fraction reaches at the most
$0.002$ (figure 2). The number ratio of helium to hydrogen is always
about 0.1 after the Big-bang.  A constant mass fraction ($Y=0.24$) of
helium is assumed throughout our calculation.

We solve equation (6) coupled with non-equilibrium rate equations for
these gaseous species by the fourth-order Runge-Kutta scheme. 
These rate equations with rate coefficients and
the heating/cooling rates are summarised in Appendix B.  In the
calculation, the time-step is adjusted to being 1/1000 of 
$-X_{\rm HI}/(dX_{\rm HI}/dt)$ for $dX_{\rm HI}/dt<0$ and 
$X_{\rm HII}/(dX_{\rm HI}/dt)$ for $dX_{\rm HI}/dt>0$ 
(see equation [B6] for $dX_{\rm HI}/dt$).
The number density of the IG dust at each
redshift is determined by equation (1), (2), and (5) depending on 
$\chi$.  The grain charge and heating/cooling rates are determined by a
standard manner which is described in \cite{ino03} 
(see also Appendix A).

The initial condition is as follows: the starting redshift is $z=3.4$,
at which it is considered that the HeII reionization occurred (e.g.,
\citealt{the02a}). The initial temperature is set to be 25,000 K, which
is the mean IGM temperature at this redshift suggested by the Lyman
$\alpha$ forest in QSO spectra \citep{sch00,the02b}. We assume an
ionization equilibrium balanced between the recombination and the
photoionization as the initial chemical abundance.  In each time step,
the calculated chemical abundance at $z<3.4$ reaches almost another
ionization equilibrium balanced among the recombination, the
photoionization, and the collisional ionization.\footnote{For a
technical reason, we did not include the collisional ionization in the
calculation of the initial condition; to avoid being 
$dX_{\rm HI}/dt=0$ for the time-step. 
Since the collisional ionization plays only a minor
role, the calculated chemical abundance is different slightly from the
abundance in the recombination--photoionization equilibrium.}

The background radiation is required to calculate the IGM thermal
history. We assume a single power-law background radiation; its mean
intensity at a frequency of $\nu$ is 
$J_\nu=J_{\rm L}(\nu/\nu_{\rm L})^{-\alpha}$,
where $J_{\rm L}$ and $\nu_{\rm L}$ are the mean intensity and the
frequency at the hydrogen Lyman limit. We also assume the spectral index
$\alpha$ to be constant, but the intensity $J_{\rm L}$ evolves along
the redshift. In figure 5, such an evolution is displayed. The
data are taken from \cite{sco02} who investigate the QSO proximity
effect on the number density of the Lyman $\alpha$ forest in spectra of
QSOs at $z=0$--5, and estimate the Lyman limit intensity of the
background radiation in the redshift range. Here, we use a fitting
formula as 
\begin{equation}
 \frac{J_{\rm L}}{\rm erg\,s^{-1}\,cm^{-2}\,sr^{-1}\,Hz^{-1}}
  \approx 2.5\times10^{-23} (1+z)^{2.5}
\end{equation}
for $0\la z \la 4$, which is shown in figure 5 as the solid line.  The
background radiation at $z\la 3$ is likely to be dominated by QSOs.
Hence, we consider two cases of $\alpha=1$ and 2. Such values of
$\alpha$ are consistent with the QSO dominated background radiation
\citep{haa96,zhe97}. In appendix A2, we show that the results with
$\alpha=1$ single power-law background spectrum should be consistent
quantitatively with those with a more realistic spectrum like
\cite{haa96}.

\begin{figure}
 \includegraphics[height=8.0cm,clip] {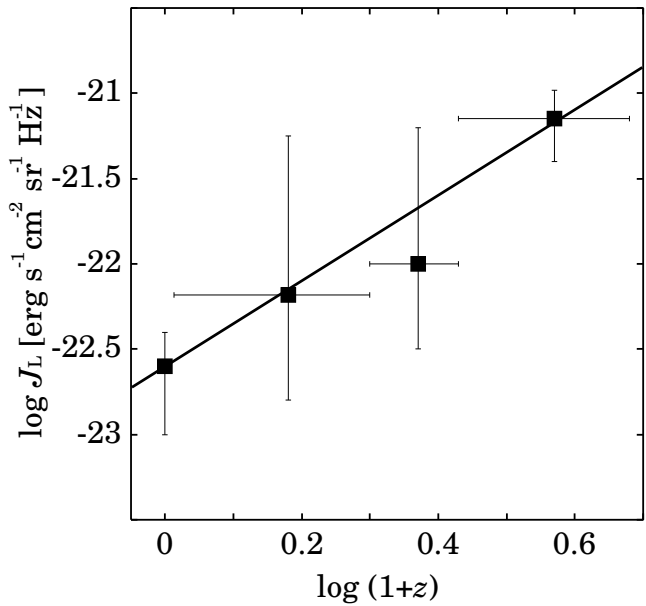}
 \caption{Redshift evolution of ionizing background intensity. Data
 points with error-bars are taken from \citet{sco02}. The solid line is
 an analytical fit described in equation (7).}
\end{figure}

In figure 6, we show some examples of the IGM thermal history, assuming
0.1 \micron\ IG dust and the spectral index of $\alpha=1$. The panel (a)
shows the silicate and the high SFH case, and the panel (b) shows the
graphite and the low SFH case. In each panel, 
six cases of $\chi$ are depicted as the solid curves. By
definition of $\chi$ (eq.[3]), $\chi=0$ means no IG dust and $\chi=1$
means that all of metal is condensed into the IG dust.
The observational data are taken from \cite{sch00}. They observe the
Ly$\alpha$ clouds with the column density of $10^{13-15}$
cm$^{-2}$ (i.e., slightly over density regions), and convert the
temperature of the clouds into that at the mean density of the IGM by
using the equation of state of the IGM. Thus, we can compare both
thermal histories directly. In the next subsection, such a comparison is
presented quantitatively.

\begin{figure}
 \includegraphics[height=8.0cm,clip] {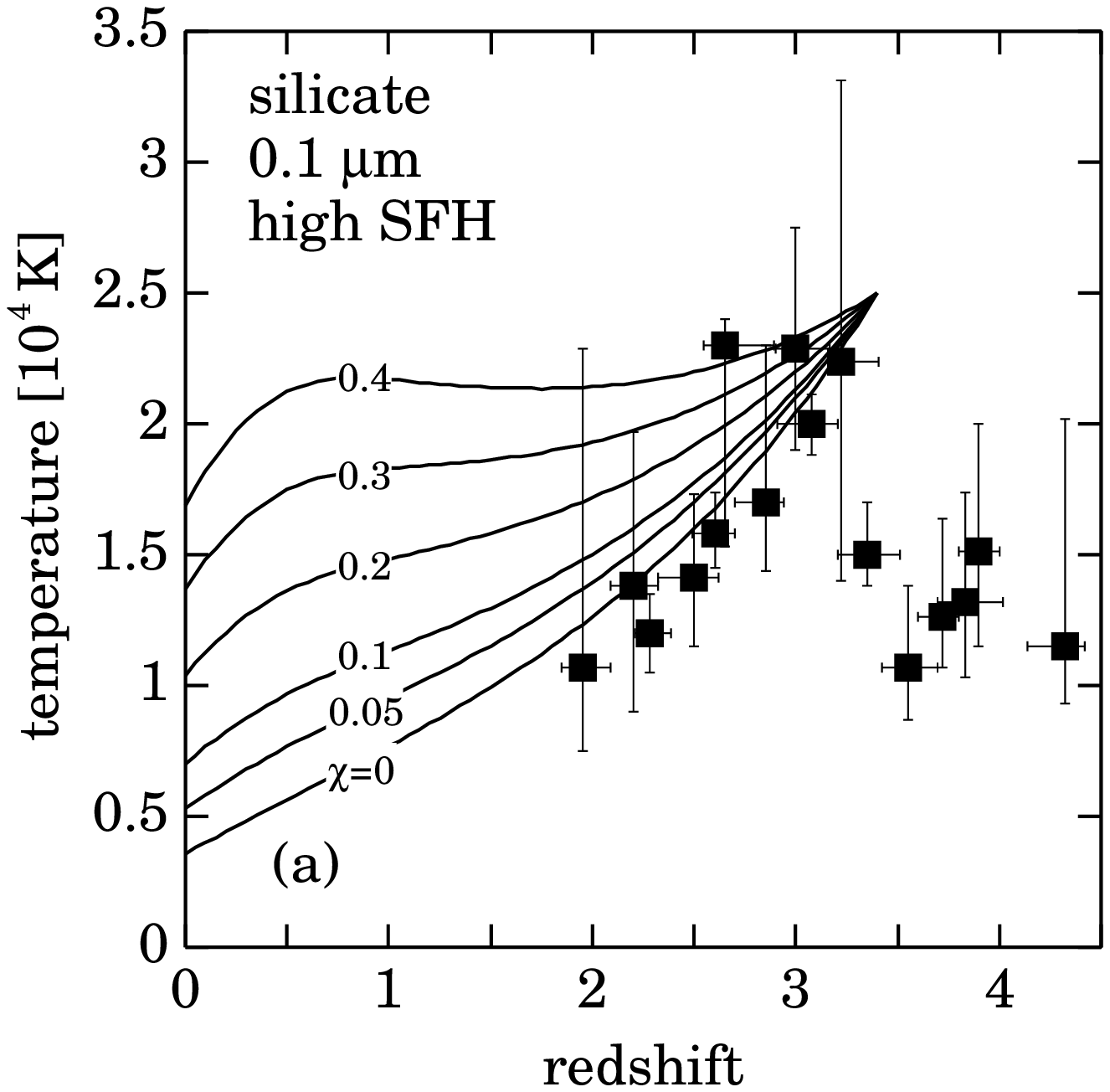}
 \includegraphics[height=8.0cm,clip] {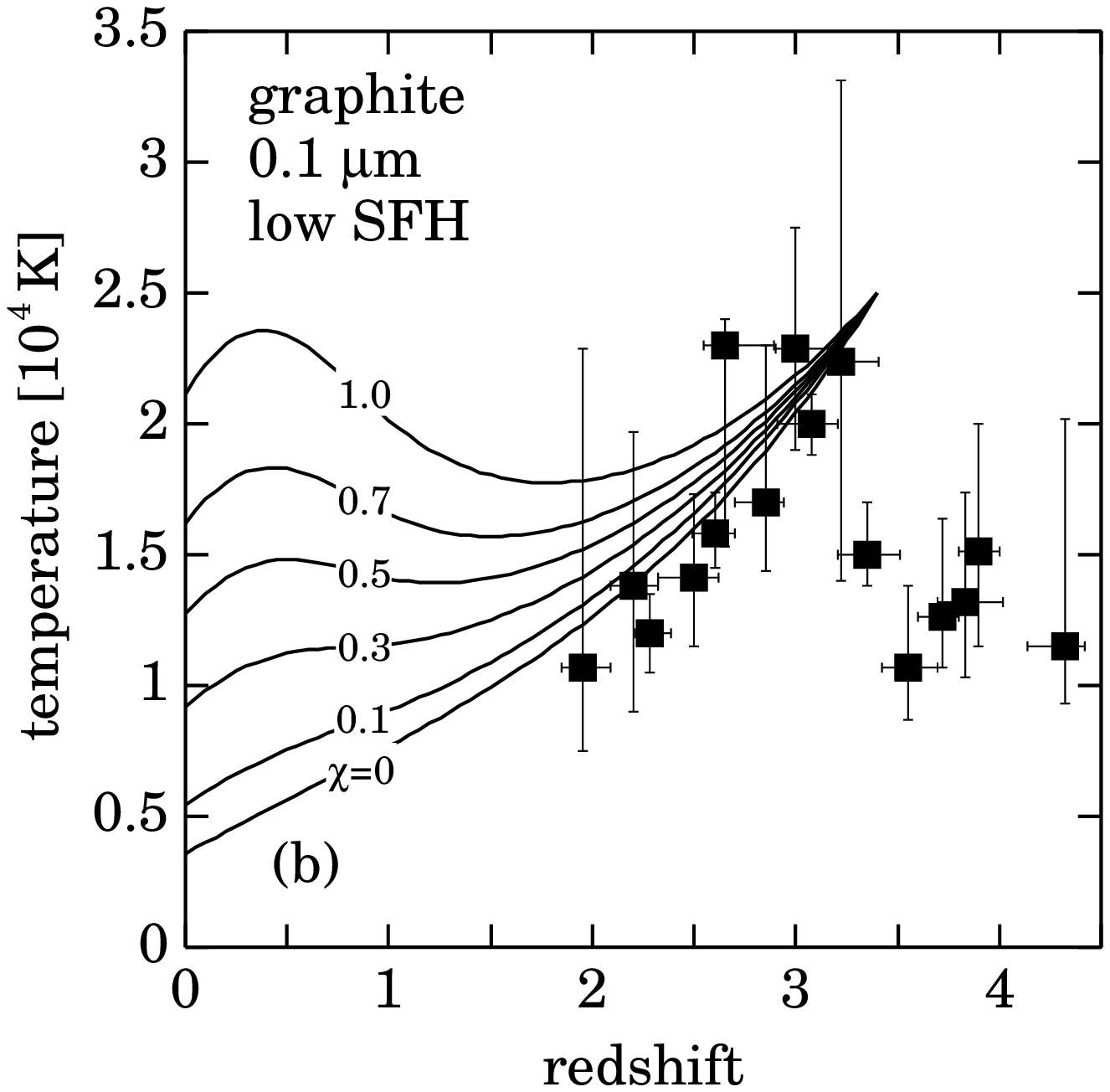}
 \caption{Examples of the IGM thermal history: (a) silicate for high SFH
 and (b) graphite for low SFH cases. The grain size is assumed to be 0.1
 \micron. Six solid curves in each panel are the thermal histories
 assumed a value of $\chi$, the mass ratio of the IG dust to
 the total metal, indicated on each curve. The data points are taken
 from \citet{sch00}. The spectral index of the background radiation is
 assumed to be unity.}
\end{figure}

\subsection{Constraint for $\chi$ from thermal history of IGM}

Once theoretical histories of IGM temperature are obtained, the amount
of IG dust is constrained from the comparison of the theoretical
temperature with observational one. Hence, we compare our theoretical
thermal histories with 10 observational points at the range of 
$1.5 < z < 3.4$ of \cite{sch00}. We reject a case of too much $\chi$ by
the least squares method. The rejection criterion is the significance
level less than 30\% in $\chi^2$-test. The obtained upper bound of
$\chi$ in the high SFH case is shown in figure 4 as the solid curves.  
In this figure, the results from distant SNe Ia are overlaid as the dotted
and dashed curves. Moreover, the rejection areas from the thermal
history and distant SNe Ia are distinguished by the thick and thin
shadings, respectively.

Combining constraints from the thermal history with those from distant
SNe Ia, we obtain the rejected range of $\chi$ as a function of the IG
grain size, which are summarised in Table 1. 
We find a rough upper bound of $\chi$ as 0.1 with a factor of a
few uncertainty, except for a very large ($\sim 1$ \micron) case and a
medium-size ($\sim 100$ \AA) silicate of $\alpha=2$.

\begin{table}
 \centering
 \begin{minipage}{140mm}
  \caption{Upper bounds of $\chi$}
  \begin{tabular}{ccccc}
   \hline
   silicate & & \\
   $\alpha=1$ & low SFH & & high SFH & \\
   10 \AA & 0.093 & TH\footnote{By thermal history.} & 0.022 & TH \\
   100 \AA & 0.22 & TH & 0.051 & TH \\
   0.1 \micron & 0.096 & SNe\footnote{By SNe Ia observations} & 0.064 & SNe\\
   1 \micron & 1.0 & def\footnote{By definition} & 0.97 & SNe\\
   $\alpha=2$ & low SFH & & high SFH & \\
   10 \AA & 0.55 & TH & 0.13 & TH \\
   100 \AA & 1.0 & def & 0.34 & TH \\
   0.1 \micron & 0.096 & SNe & 0.064 & SNe\\
   1 \micron & 1.0 & def & 0.97 & SNe\\
   \hline
   graphite & & \\
   $\alpha=1$ & low SFH & & high SFH & \\
   10 \AA & 0.17 & SNe & 0.041 & TH \\
   100 \AA & 0.15 & SNe & 0.076 & TH \\
   0.1 \micron & 0.11 & SNe & 0.071 & SNe\\
   1 \micron & 1.0 & def & 0.95 & SNe\\
   $\alpha=2$ & low SFH & & high SFH & \\
   10 \AA & 0.17 & SNe & 0.11 & SNe \\
   100 \AA & 0.15 & SNe & 0.10 & SNe \\
   0.1 \micron & 0.11 & SNe & 0.071 & SNe\\
   1 \micron & 1.0 & def & 0.95 & SNe\\
   \hline
  \end{tabular}
 \end{minipage}
\end{table}

The upper bound of $\chi$ from the thermal history has a
positive dependence of grain size. This corresponds to the fact that the
dust heating rate has a negative dependence of grain size (see figure
A3). Especially, for small silicate grain, we obtain a strict upper
bound of $\chi$ from the thermal history, whereas observations of
distant SNe Ia cannot provide any constraints.

We notice here that the upper bound of $\chi$ for small silicate is
smaller than that for small graphite. This is why a small ($a \la 0.1
\micron$) silicate has a larger efficiency factor for absorption than
that of graphite in the UV--X-ray regime. Hence the grain potential,
mean photoelectron energy, and heating rate of small silicate are larger
than those of graphite (figures A1--A3).  Moreover, we find a positive
dependence of the spectral index $\alpha$ against the obtained upper
bounds of $\chi$, which is due to the negative dependence of $\alpha$
against the dust heating rate (figure A3).

\section{Discussion}

We have obtained upper bound of $\chi$ as a function of grain size from
observations of distant SNe Ia and comparison of theoretical IGM thermal
history with observational one.  Here we discuss what our results imply.

\subsection{Allowable amount of IG dust}

\begin{figure}
 \includegraphics[height=8.0cm,clip] {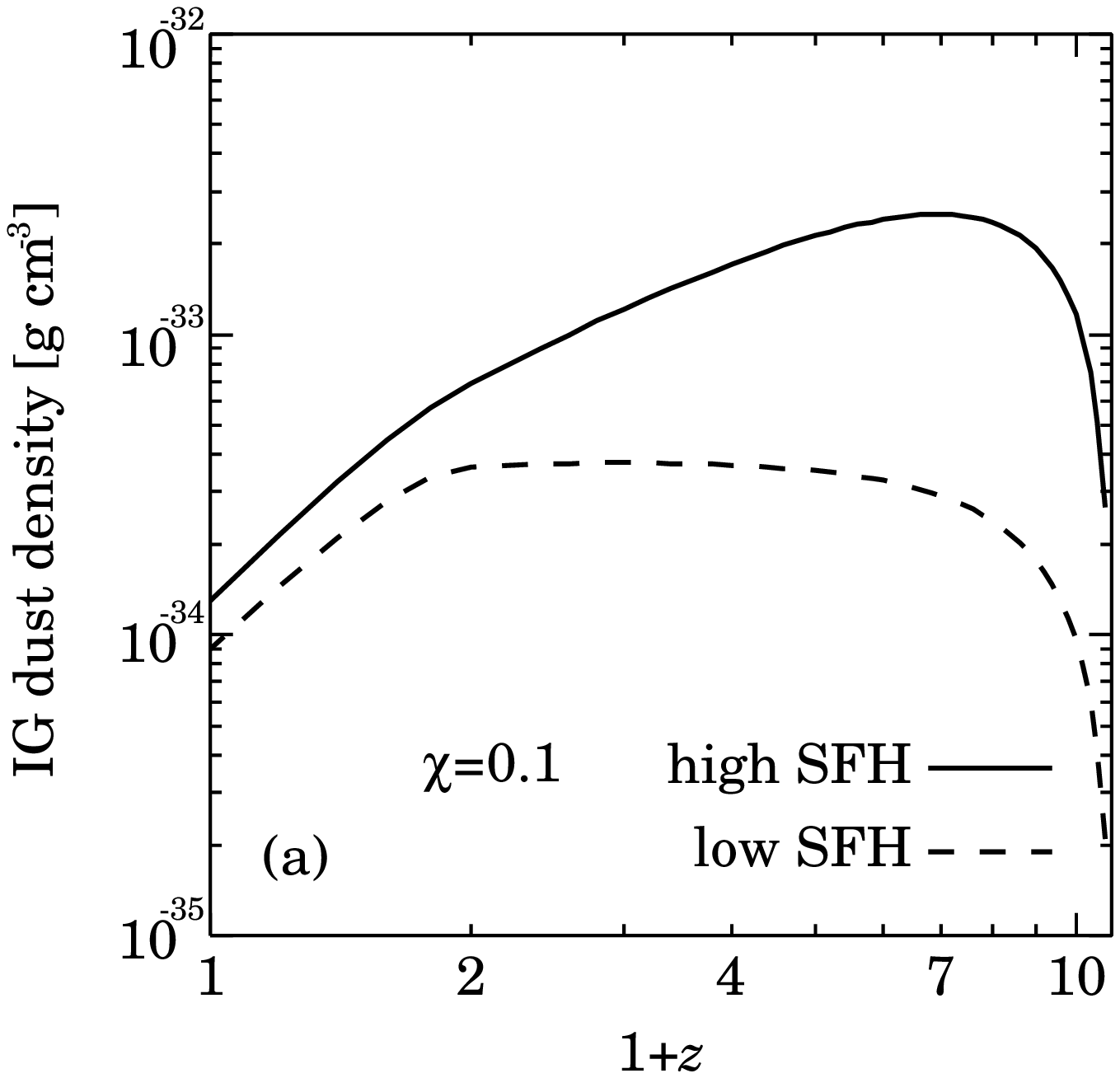}
 \includegraphics[height=8.0cm,clip] {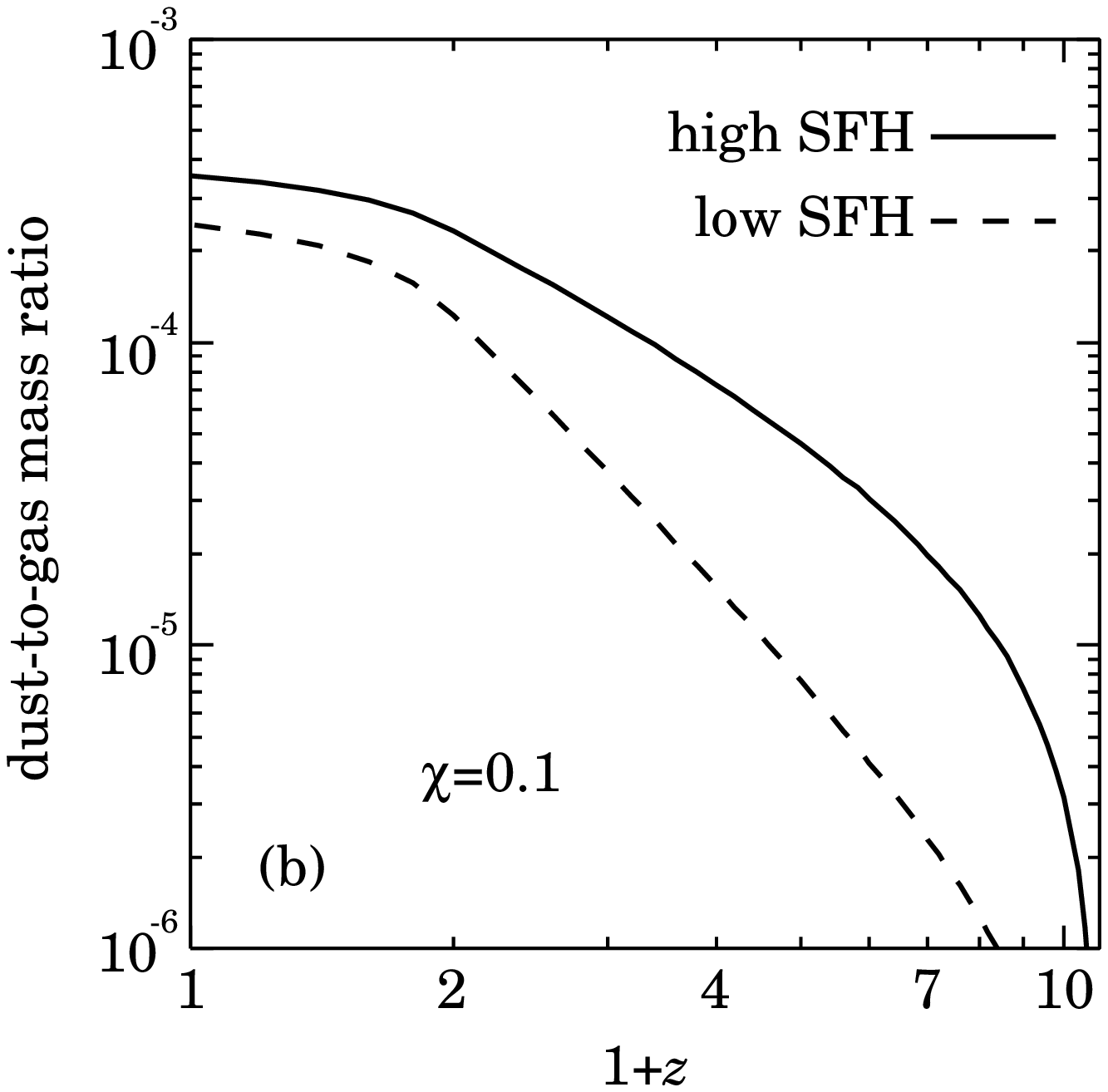}
 \caption{Maximum (a) IG dust density and (b) dust-to-gas mass
 ratio in the IGM as a function of redshift. In each panel, the solid
 and dashed curves are the cases of high and low star formation
 histories, respectively. The assumed $\chi$, the mass ratio of the IG
 dust to the total metal in the universe, is 0.1.}
\end{figure}

How much dust can exist in the IGM?  Once assuming a value of $\chi$, we
obtain the IG dust density via equation (5). In figure 7, we show the
upper bounds of the IG dust mass density and ${\cal D}^{\rm IGM}$ as a
function of redshift. The solid and dashed curves are the cases of high 
and low SFHs, respectively. The assumed upper bound of $\chi$ is 0.1
for both cases of SFHs.  The uncertainty of this value of $\chi$ is
about a factor of a few as long as the IG grain size is smaller than 1
\micron\ and the background spectral index $\alpha=1$ (see also Table 1).  
We find that the upper bound of the local ($z\sim 0$) universe is
determined well; the local IG dust density is less than about $10^{-34}$
g cm$^{-3}$, or equivalently the dust-to-gas ratio is less than about
$3\times 10^{-4}$ which is about 1/20 of the Galactic value. 
Along the redshift,
the allowed dust density increases and the dust-to-gas ratio
decreases. The increasing/decreasing rates change at $z\sim 1$ at which 
the $(1+z)$ dependence of the assumed SFH changes (eq.[1]).  Taking into
account a factor of 2 uncertainty of the upper bound of $\chi$, we
obtain the maximum IG dust density as 
\begin{equation}
 \frac{\rho_{\rm d,max}^{\rm IGM}(z)}{\rm g\,cm^{-3}} = 
  (2-8)\times 10^{-34}
  \cases{
  \left(\frac{1+z}{2}\right)^{2.3} & (for $z\leq 1$) \cr
  \left(\frac{1+z}{2}\right)^{0-1.5} & (for $z>1$) \cr
  }\,, 
\end{equation}
or the maximum dust-to-gas mass ratio as 
\begin{equation}
 {\cal D}_{\rm max}^{\rm IGM}(z) = 
  (1-4)\times 10^{-4}
  \cases{
  \left(\frac{1+z}{2}\right)^{-0.7} & (for $z\leq 1$) \cr
  \left(\frac{1+z}{2}\right)^{-(1.5-3)} & (for $z>1$) \cr
  }\,.
\end{equation}
The dust-to-gas ratio at $z\sim 3$ is consistent with the previous
result by \cite{ino03}.

\subsection{IG extinction and reddening}

\begin{figure}
 \includegraphics[height=8.0cm,clip] {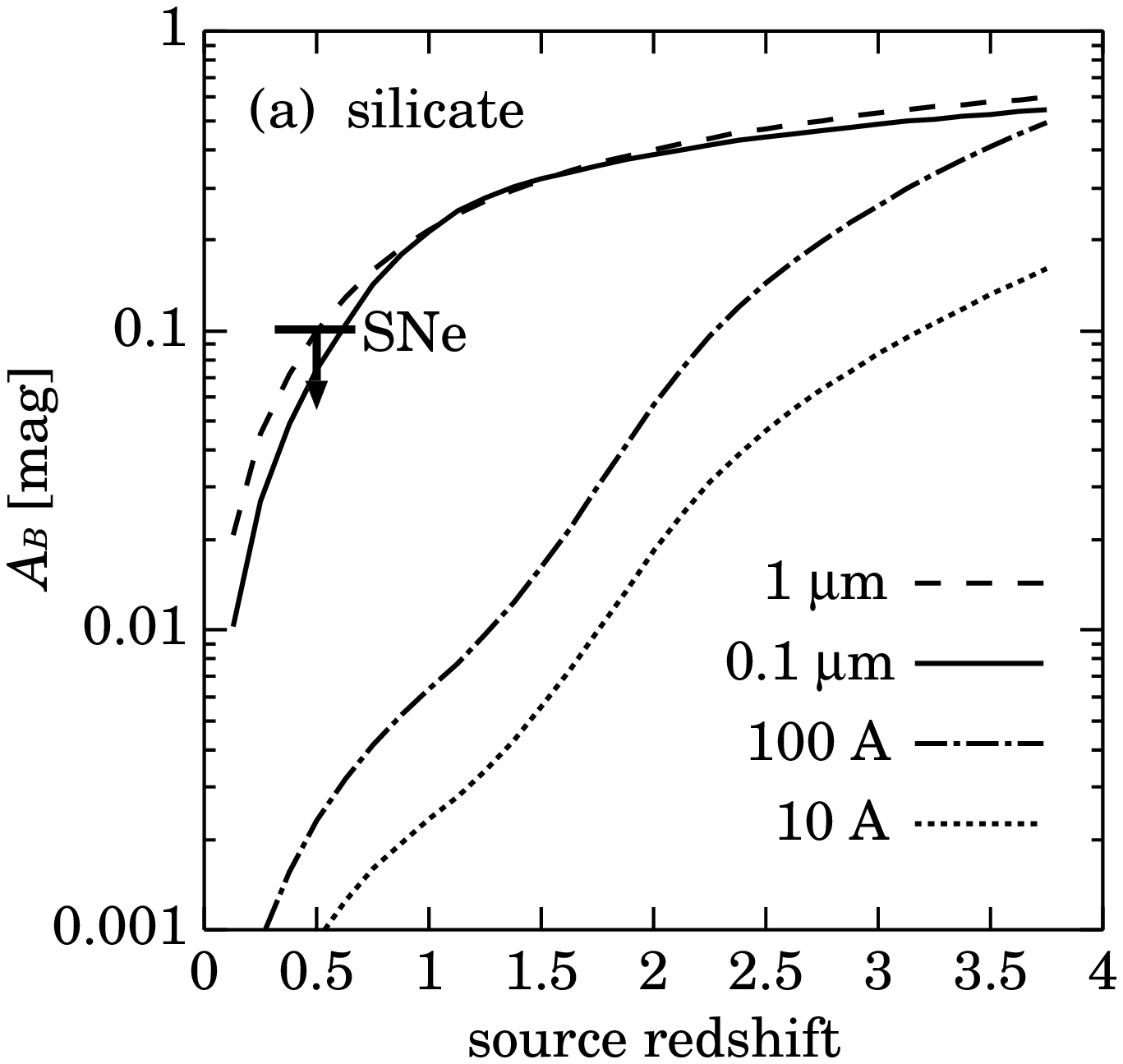}
 \includegraphics[height=8.0cm,clip] {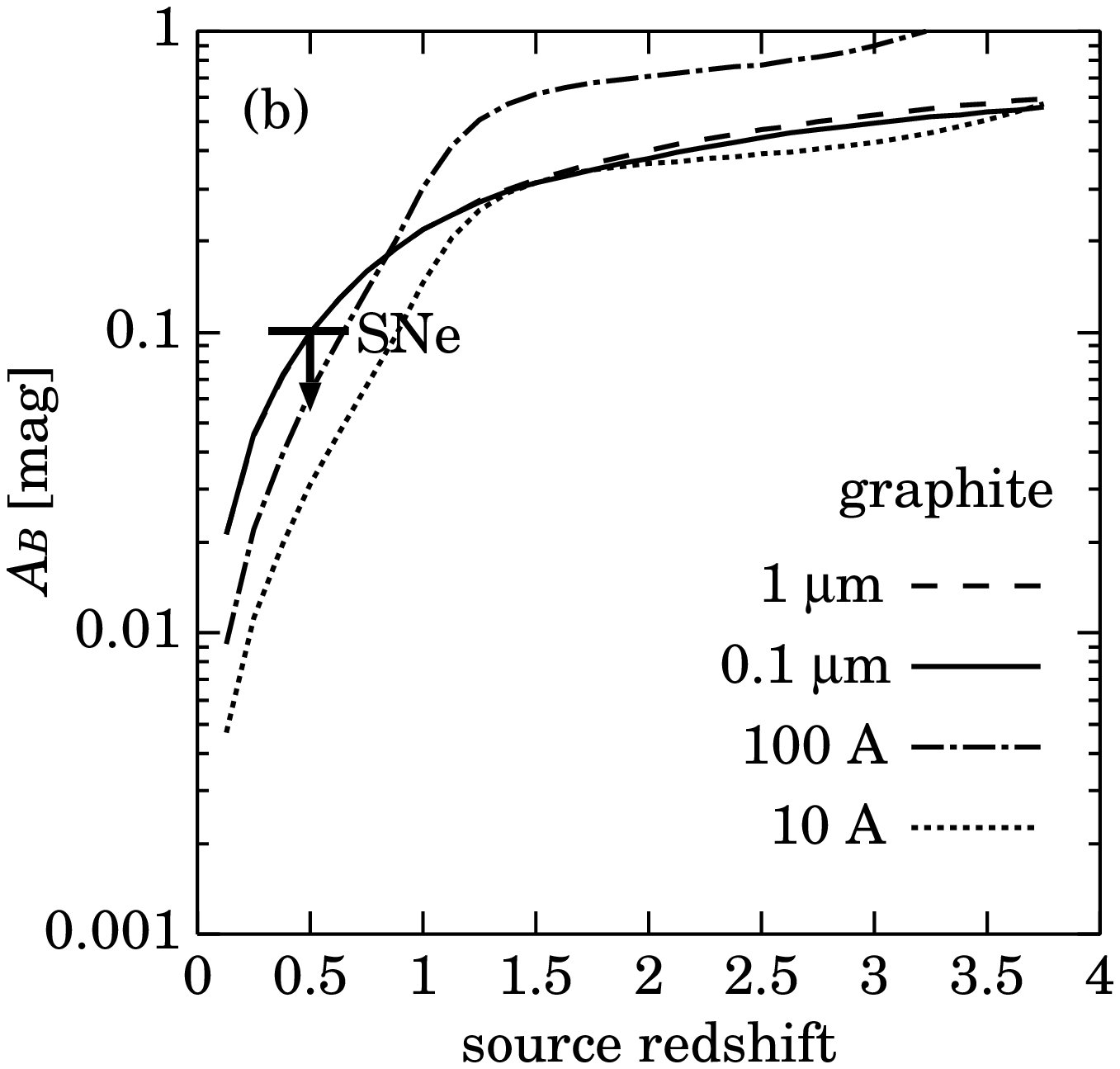}
 \includegraphics[height=8.0cm,clip] {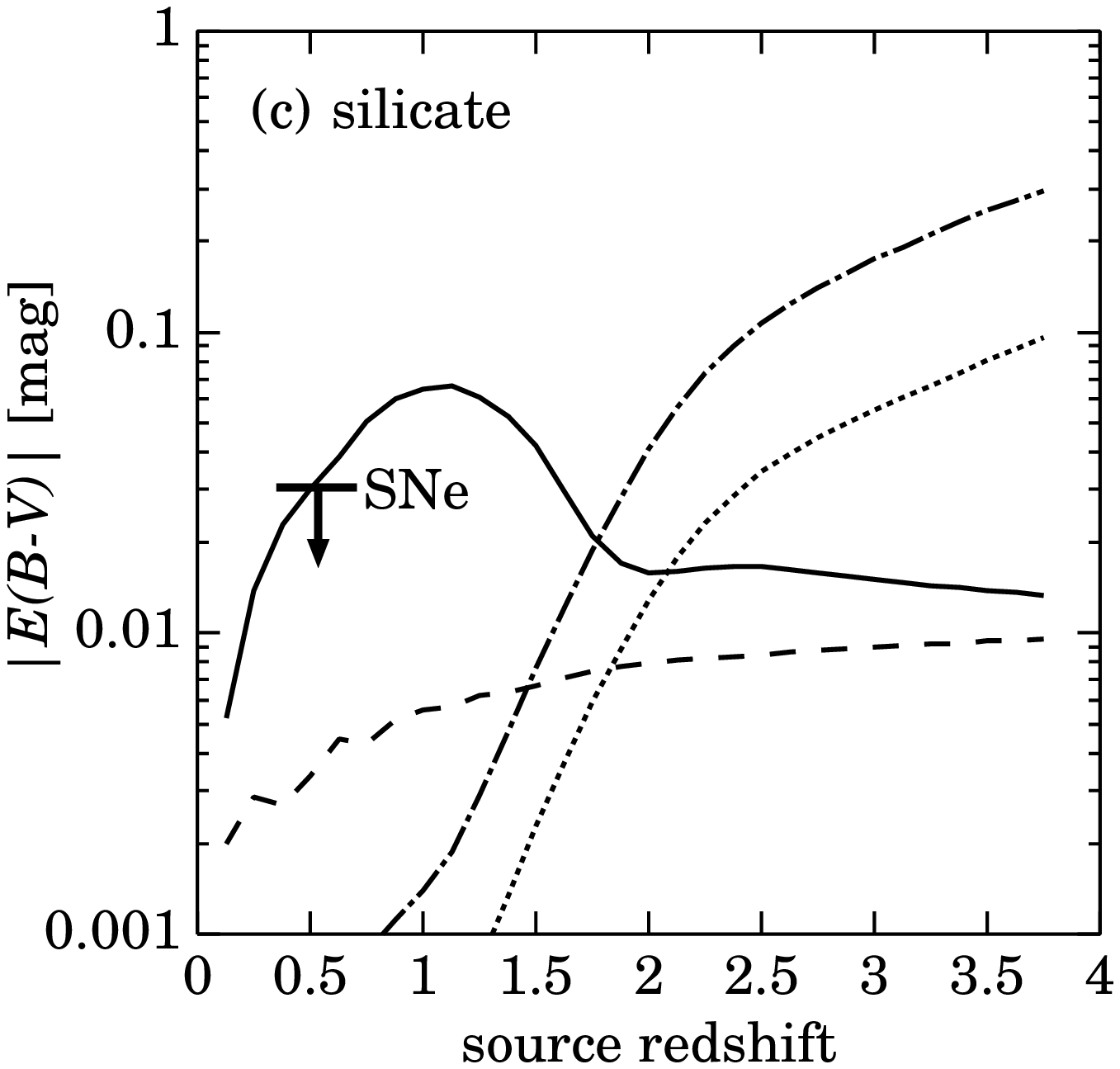}
 \includegraphics[height=8.0cm,clip] {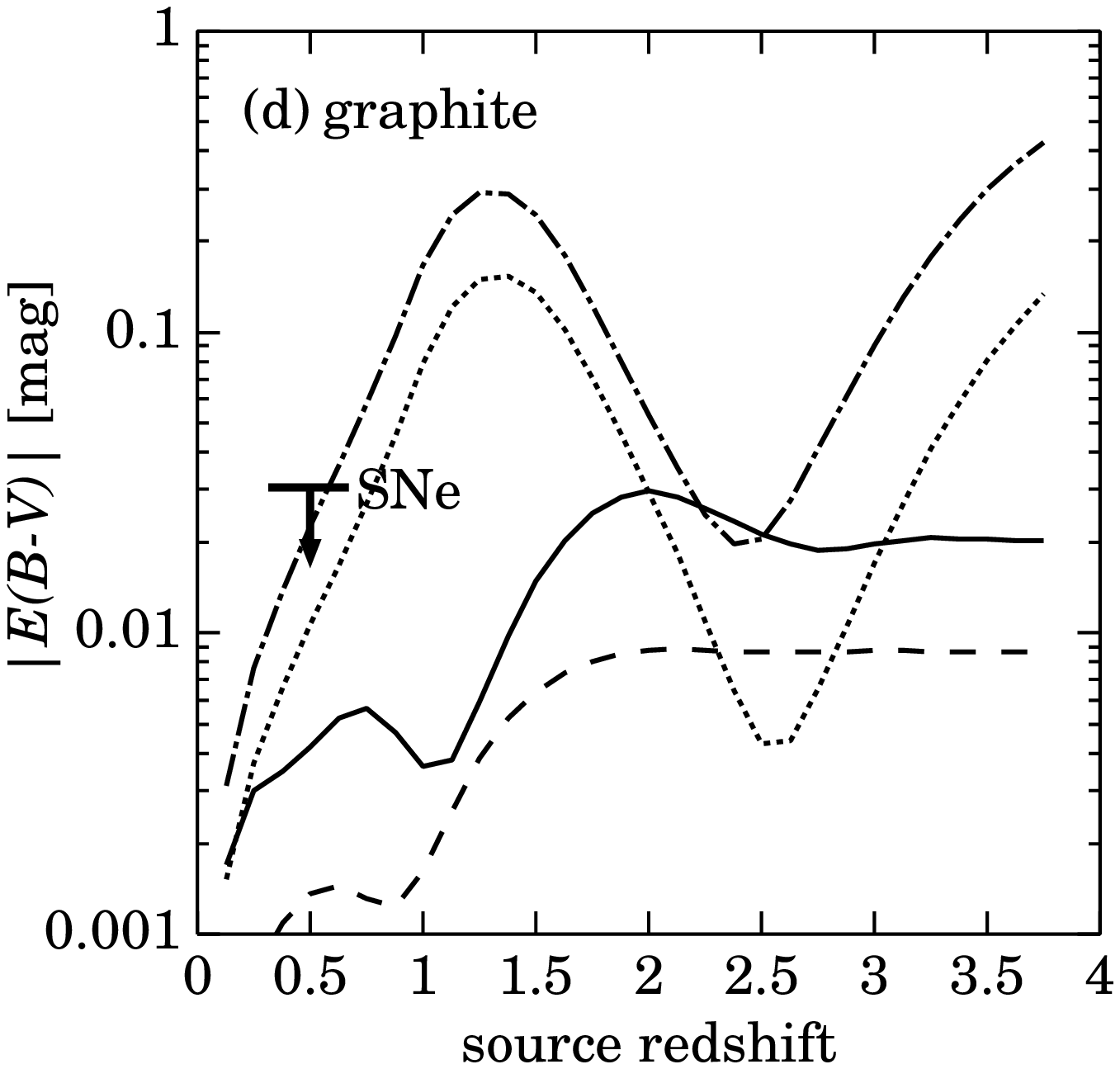} \caption{Maximum
 intergalactic extinction and reddening against a source at a redshift
 in the case of high star formation history;
 observer's $B$-band extinction of (a) silicate case and (b) graphite
 case, and colour excesses between $B$ and $V$-bands in observer's rest
 frame of (c) silicate case and (d) graphite case. Assumed $\chi$ values
 are summarised in Table 1. For dotted, dot-dashed, solid, and dashed
 curves, we assumed the grain size of 10 \AA, 100 \AA, 0.1 \micron, and
 1 \micron, respectively. The spectral index of the background radiation
 is assumed to be unity. The upper bounds from observations of distant
 SNe Ia are shown as the downward arrow in each panel.  The vertical axes
 of panels (c) and (d) are the absolute value of the colour
 excesses. Actually, colour excesses of 1 \micron\ for silicate and of
 0.1 and 1 \micron\ for graphite are negative.}
\end{figure}

In figure 8, we show the upper bounds of IG extinction and reddening
expected from the upper bounds of $\chi$ in the case of the high SFH and
$\alpha=1$ summarised in Table 1. 
Four cases of grain size as 10 \AA, 100 \AA, 0.1
\micron, and 1 \micron\ are indicated by dotted, dot-dashed, solid, and
dashed curves, respectively. The constraints from SNe Ia observations of 
$A_B \leq 0.1$ mag and $|E(B-V)| \leq 0.03$ mag at $z=0.5$ are also
shown as the downward arrow in each panel.  We note that the vertical
axes of panels (c) and (d) are the absolute value of colour
excess. Indeed, colour excesses of 1 \micron\ for silicate, and of 0.1
\micron\ and 1 \micron\ for graphite are negative.

We find that the upper bound of the IG extinction is $\sim 0.2$ mag for
a source at $z=1$ from panels (a) and (b) of figure 8. This value
agrees well with the result from SDSS quasars data by \cite{mor03}. For
$z\ga1$ objects, the upper bound of the IG extinction becomes $\sim 0.5$
mag, and as an extreme case, we cannot reject the possibility of 1 mag
IG extinction for a source at $z\sim 3$.

Interestingly, we can investigate the nature of the IG dust by using the
IG reddening. For $z\ga 1$ sources, the expected absolute value of
colour excess by the IG grain larger than $\sim 0.1$ \micron\ is very
small, at the most $\sim 0.05$ mag, whereas that by a smaller grain can
reach 0.1 mag or more. Thus, it may be possible to determine a typical
size of the IG grain from observations of colour excess against a source
at $z\ga1$; 
the detection of $\ga 0.1$ mag colour excess for such a source proves
the existence of small ($\la 100$ \AA) IG grains. If the IG dust is
dominated by such small grains, the composition of the IG dust can be
found. The small graphite grains show a prominent absorption feature at
2175 \AA. Thus, we expect a local minimum of colour
excess for a source at $z\simeq 2.5$ as shown in panel (d) of figure 8. 
Therefore, if we detect such a change of colour excess along the
redshift, we can conclude that many small graphite grains exist in the
IGM.  

Observations to detect the IG extinction and reddening are very
challenging but strongly encouraging. 
High-$z$ gamma-ray bursts can be good background light sources for
such observations \citep{per00}.

\subsection{Ejection efficiency of dust from galaxies}

The dust-to-metal mass ratio in the Galaxy, $\delta_{\rm MW}$, is
0.3--0.5 \citep[e.g.,][]{spi78,whi03}.  We may consider the
dust-to-metal ratio in the IGM, $\delta_{\rm IGM}$ is equal to
$\delta_{\rm MW}$, if $\delta_{\rm MW}$ is typical for all galaxies, and
metal and dust are ejected together from galaxies to the IGM keeping
$\delta_{\rm MW}$. Is this compatible with the obtained upper bound of
$\chi \la 0.1$? 

A fraction of metal produced by stars in galaxies exists out of
galaxies. This metal escape fraction is defined as $f_{\rm Z,esc}$.
While $f_{\rm Z,esc}$ is still uncertain, an estimate of it is 50--75\%
\citep[][and references therein, see also our figure 2]{agu99}.  
We shall define another parameter of the IGM metallicity; $Z_{\rm IGM}$.  
Because of $\chi = {\cal D}^{\rm IGM}/Z$ and $f_{\rm Z,esc} = Z_{\rm
IGM}/Z$, where $Z$ is the total cosmic metallicity, we find 
$\delta_{\rm IGM}= {\cal D}^{\rm IGM}/Z_{\rm IGM} = \chi /f_{\rm Z,esc}$.
Unless $f_{\rm Z,esc}$ is less than $\sim 0.5$, 
$\delta_{\rm IGM}$ is estimated to be smaller than 0.2 if $\chi \la 0.1$. 
Therefore, our result of $\chi \la 0.1$ with $f_{\rm Z,esc} \sim 0.5$
may indicate that $\delta_{\rm IGM} < \delta_{\rm MW}$. 

If that is true, we have to consider some mechanisms to reduce 
$\delta_{\rm IGM}$ during the dust transfer.  For example, dust
destruction during the transfer from galaxies to the IGM \citep{agu99},
and/or different ejection efficiencies between metal and dust. Time
evolution of the dust-to-metal ratio in galaxies may be also
important. As shown by \cite{i03}, the dust-to-metal ratio in younger
galaxies (i.e., higher-$z$ galaxies) may be much smaller ($\sim 70$\%
off) than the present value of the Galaxy.  In the case, a time-averaged
$\delta_{\rm MW}$ can become smaller than the current
$\delta_{\rm MW}$ adopted above, so that our constraint of
$\chi$ may be cleared.  In any case, we cannot obtain a rigid
quantitative conclusion at the moment, because uncertainties are still
large. Further studies of this issue are very interesting.

\subsection{IGM temperature at low redshift}

As shown in Nath et al.~(1999) and Appendix A, the dust photoelectric
heating becomes more efficient for a lower gas density. Although the
background intensity decreases along the redshift (figure 5), the
decrement of gas density is more efficient than the decrement of the
background intensity, so that the importance of the dust heating
increases for a lower redshift. While we obtained constraints of the
amount of the IG dust from the IGM temperature at $z\simeq2$--3 in
section 4, the IGM temperature at a lower redshift of $z\la1$
provides us with a further constraint of the IG dust. Therefore, to
measure the IGM temperature at $z\la1$ is very interesting.

Here, we demonstrate how much temperature is allowed by our upper bounds
of $\chi$. Figure 9 shows the IGM thermal histories assumed the upper
bounds of $\chi$ in Table 1 for the case of background
spectral index $\alpha=1$ and high SFH. The dotted, dot-dashed, and
solid curves are the case of no IG dust, 100 \AA\ silicate, and 0.1
\micron\ silicate, respectively. We can make a very similar figure for
graphite case.  The temperatures shown in the figure are
upper bounds, which is denoted as $T_{\rm IGM,up}$.

\begin{figure}
 \includegraphics[height=8.0cm,clip] {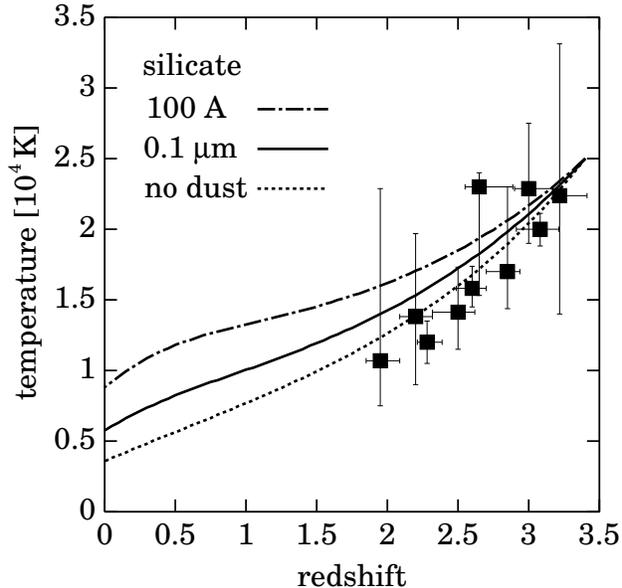}
 \caption{Maximum IGM temperatures corresponding to the upper bounds of
 $\chi$ in table 1 for the case of silicate, background spectral
 index of unity, and high SFH. The dotted, dot-dashed, and solid curves
 are the case of no IG dust, 100 \AA, and 0.1 \micron, respectively.
 The data points are taken from \citet{sch00}.}
\end{figure}

After checking all cases listed in table 1, 
we find that for a smaller ($\la 100$ \AA) grain case, except for
graphite of $\alpha=2$, $T_{\rm IGM,up}$ at $z\la1$ is still much higher
than 10,000 K. On the other hand, for a larger ($\ga 0.1$ \micron) case,
except for 1 \micron\ graphite with high SFH and $\alpha=1$, $T_{\rm
IGM,up}$ at $z\la1$ becomes lower than 10,000 K as well as no IG dust
case.  Therefore, we may conclude that IG grains are small if
temperature higher than 10,000 K is observed at $z\la1$. Conversely, a
lower IGM temperature at $z\la1$ provides us with a very strict
constraint against small IG dust.

\section{Conclusion}

We investigate the amount of the IG dust allowed by current observations
of distant SNe Ia and temperature of the IGM.  The allowed amount of the IG
dust is described as the upper bound of $\chi$, the mass ratio of the IG
dust to the total metal mass in the Universe. To specify $\chi$, two
models of cosmic history of metal production rate are assumed. That is,
we have assumed two cosmic star formation histories expected from the
recent observation of the high redshift objects. 
Our conclusions are as follows:

(1) Combining constraints from the IGM thermal history with those from
distant SNe Ia observations, we obtain the upper bounds of $\chi$ as a
function of grain size in the IGM; roughly $\chi \la 0.1$ for 10 \AA\
$\la a\la 0.1$ \micron, and $\chi\la 0.1(a/0.1\,\micron)$ for 0.1
\micron\ $\la a \la 1$ \micron.

(2) The upper bound of $\chi\sim0.1$ corresponds to the upper bound of
the IG dust density; the density increases from $\sim10^{-34}$ g
cm$^{-3}$ at $z=0$ to $\sim10^{-33}$ g cm$^{-3}$ at $z\sim1$, and keeps
a constant value or slowly increases toward higher redshift.

(3) The expected IG extinction against a source at $z\sim 1$ is less
than $\sim 0.2$ mag at the observer's $B$-band. For higher redshift
sources, we cannot reject the possibility of 1 mag extinction by the IG
dust at the observer's $B$-band.

(4) Observations of colour excess against a source at $z\ga1$ provides
us with information useful to constrain the nature of the IG dust. If we
detect $\sim 0.1$ mag colour excess between the observer's $B$ and
$V$-bands, a typical size of the IG dust is $\la 100$ \AA.  Moreover, if
there are many graphite grains of $a \la 100$ \AA\ in the IGM, we find a
local minimum of the colour excess of a source at $z\sim2.5$
corresponding to 2175 \AA\ absorption feature.

(5) If half of metal produced in galaxies exists in the IGM, the
obtained upper bound of $\chi\sim0.1$ means that the dust-to-metal ratio
in the IGM is smaller than the current Galactic value.  It suggests that
some mechanisms to reduce the dust-to-metal ratio in the IGM are
required. For example, dust destruction in transfer from galaxies to the
IGM, selective transport of metal from galaxies, and time evolution of
the dust-to-metal ratio in galaxies 
(i.e., a smaller value for younger galaxies).

(6) Although we obtain constrains of the IG dust from the IGM
temperature at $z\sim 2$--3, the temperature at $z\la1$ provides us with
a more strict constraint of the IG dust.  For example, the detection of
temperature higher than 10,000 K at $z\la1$ suggests that the IG dust is
dominated by small ($\la 100$ \AA) grains.

\section*{Acknowledgments}

We have appreciated a lot of advice to improve the quality of this paper
by the referee, Dr. S. Bianchi, informative comments presented by
Prof. B. Draine and Dr. J. Schaye, and discussions with Prof. A. Ferrara
and Dr. H. Hirashita.  
We are also grateful to Profs. T. Nakamura, S. Mineshige, and
I.-S. Inutsuka for their encouragement.
This work is supported by a Grant-in-Aid for the 21st Century COE
''Center for Diversity and Universality in Physics''. 
AKI is supported by the Research Fellowship of the Japan Society for
the Promotion of Science for Young Scientists.

\appendix

\section[]{Dust photoelectric heating in IGM}

The dust photoelectric effect in the IGM is summarised.  The basic
equations of the dust photoelectric effect are presented in various
places, for example, section 2 in \cite{ino03}.  More detailed
information on this effect can be found in \cite{wei01b}.

We consider spherical silicate and graphite grains.  Under a condition
suitable for the IGM (low density and intense UV radiation), these
grains have a positive electric charge, which is determined mainly by
the competition between the collisional electron capture and the
photoelectric ionization. The ion collision is negligible, while the
proton collision is included in the calculation 
in section 4 and for making the following figures. 
The charge on grains is in an equilibrium state, which is achieved
quickly ($\sim 10$--100 yr).

The input parameters to obtain the equilibrium charge are grain type,
grain size, gas density, gas temperature, radiation intensity, and
radiation spectrum. We assume the incident radiation spectrum to be a
power-law. In figure A1, we show these dependences of the equilibrium
grain potential energy normalized by the gas kinetic energy, i.e.,
$eU/k_{\rm B}T=Z_{\rm d}e^2/ak_{\rm B}T$, where $U$ is the grain
potential, $Z_{\rm d}$ is the grain charge, $a$ is the grain size, $T$
is the gas temperature, $e$ is the electron charge, and $k_{\rm B}$ is
the Boltzmann constant. We examine three cases of the power-law index of
the incident radiation, $\alpha=1$, 2, and 5, where the mean intensity
is $J_\nu $ and proportional to $\nu^{-\alpha}$.  While the spectrum of
the incident radiation is rather uncertain, it is likely to be a
power-law with index of 1--2 if the radiation is dominated by QSOs
(e.g., \citealt{haa96,zhe97}). The case of $\alpha=5$ is a reference of
a very soft background radiation. The parameter set assumed in the
calculation are noted in each panel of figure A1; 
$a_{,-5}=a/0.1\micron$, $n_{,-5}=n/10^{-5}\,{\rm cm}^{-3}$,
$T_{,4}=T/10^4\,{\rm K}$, and $J_{,-21}=J_{\nu_{\rm L}}/10^{-21}\,{\rm
erg\,s^{-1}\,cm^{-2}\,sr^{-1}\,Hz^{-1}}$, where $\nu_{\rm L}$ is the
Lyman limit frequency. These values may be suitable for the IGM at
$z\sim 3$.  

The dotted lines in figure A1 (a) and (c) indicate an upper limit of the
grain potential based on an estimate of the critical potential for the
ion field emission \citep{dra02}; $eU/k_{\rm B}T \la 3500 (T/10^4\,{\rm
K})^{-1} (a/0.1\,\micron)$.  If the grain potential exceeds this upper
limit, singly charged ions may escape one by one from the grain surface,
so that the grain is destroyed gradually. For panels (b) and (d), this
upper limit is out of the panels.  We can conclude that this process is
not so important for our interest.

In figure A2, we show a mean photoelectron energy from dust grains
normalized by the gas kinetic energy, 
$\langle E_{\rm pe} \rangle/k_{\rm B}T$, 
as a function of (a) grain size, (b) gas density, (c) gas
temperature, and (d) radiation intensity. Moreover, the ratio of the
dust photoelectric heating rate to the hydrogen photoionization heating
rate ($\Gamma_{\rm pe}/\Gamma_{\rm Hpi}$) is depicted as a function of
these quantities in figure A3. For the dust heating rate, we assume the
dust-to-gas mass ratio to be $10^{-4}$ as a nominal value, which is
about 1/100 of the Galactic dust-to-gas mass ratio.

The obtained results are roughly consistent with those of
\cite{nat99}. While some quantitative differences are seen between our
results and theirs, they may be caused by differences of the adopted
photoelectric yield, absorption efficiency factors, and radiation
spectrum.

In summary, we find that the dust photoelectric heating exceeds the
photoionization heating for a case of a smaller grain size, a lower gas
density, a higher temperature, a more intense radiation, or a harder
radiation spectrum. Furthermore, we find that silicate and graphite
grains show similar results, and all of $eU/k_{\rm B}T$, $\langle E_{\rm
pe} \rangle/k_{\rm B}T$, and $\Gamma_{\rm pe}/\Gamma_{\rm Hpi}$ show a
power-law like dependence of grain size, gas density, gas temperature,
and radiation intensity. These behaviors will be derived analytically
with some approximations in the next subsection.

\begin{figure}
 \includegraphics[height=8.0cm,clip] {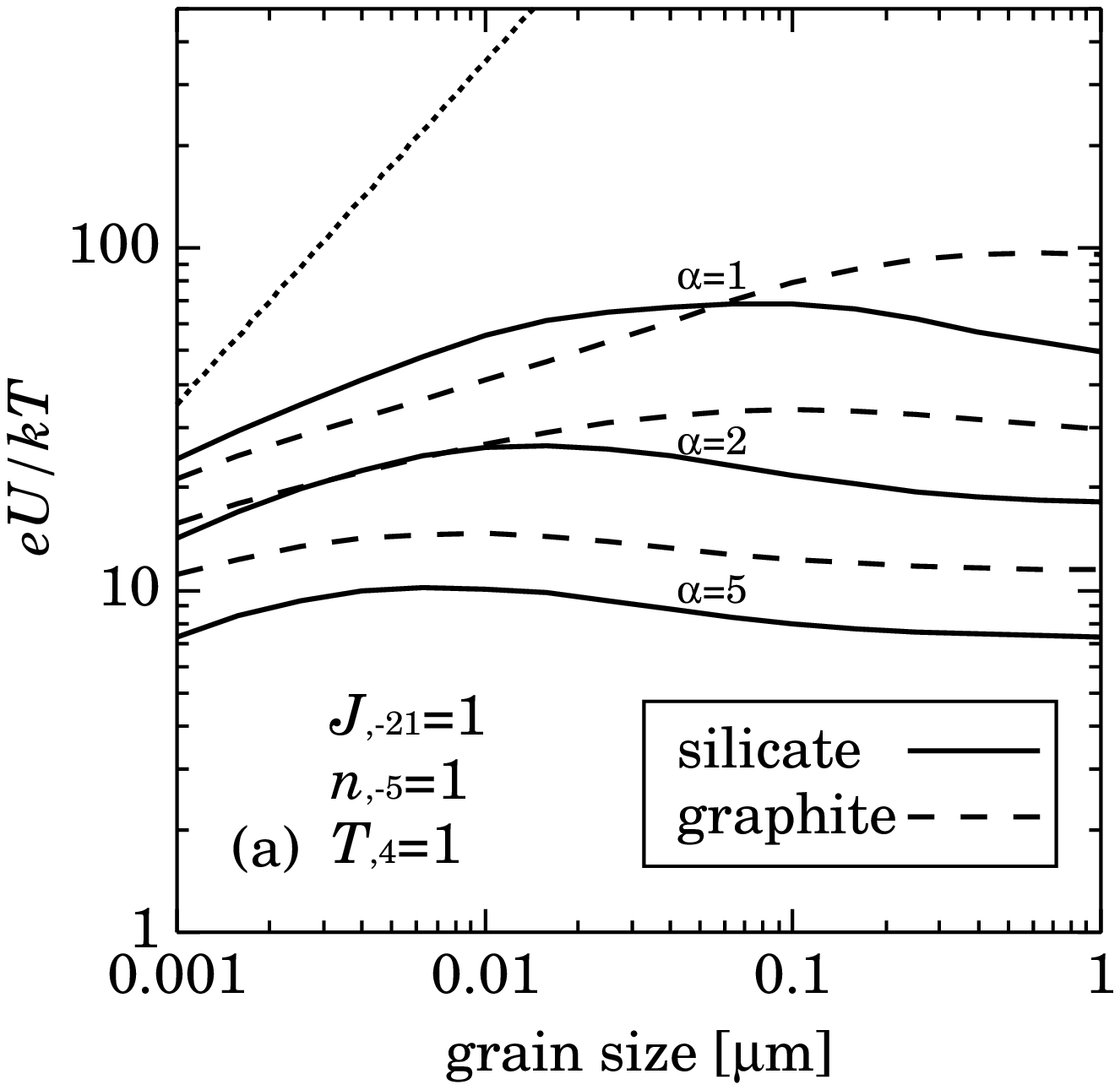}
 \includegraphics[height=8.0cm,clip] {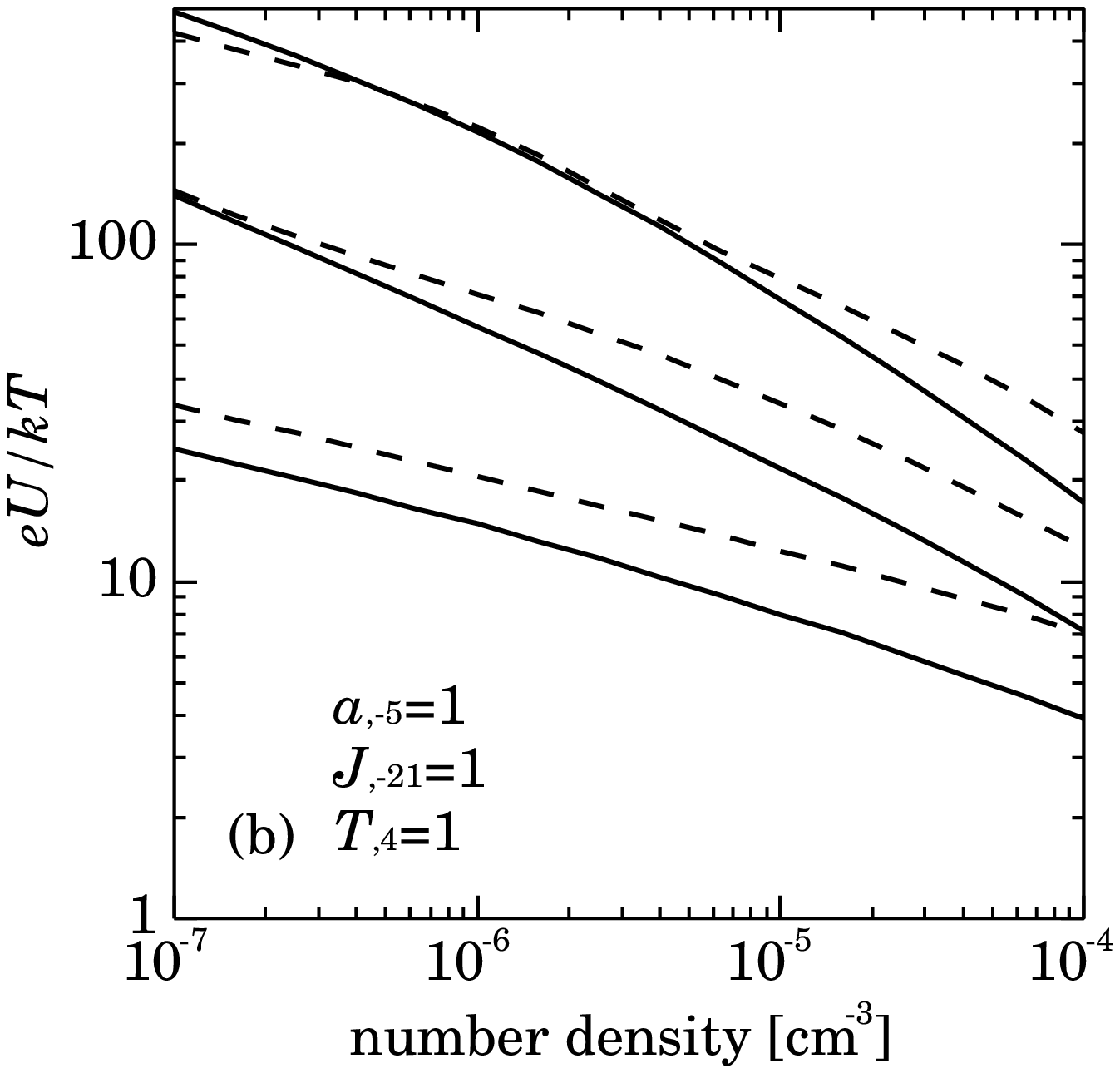}
 \includegraphics[height=8.0cm,clip] {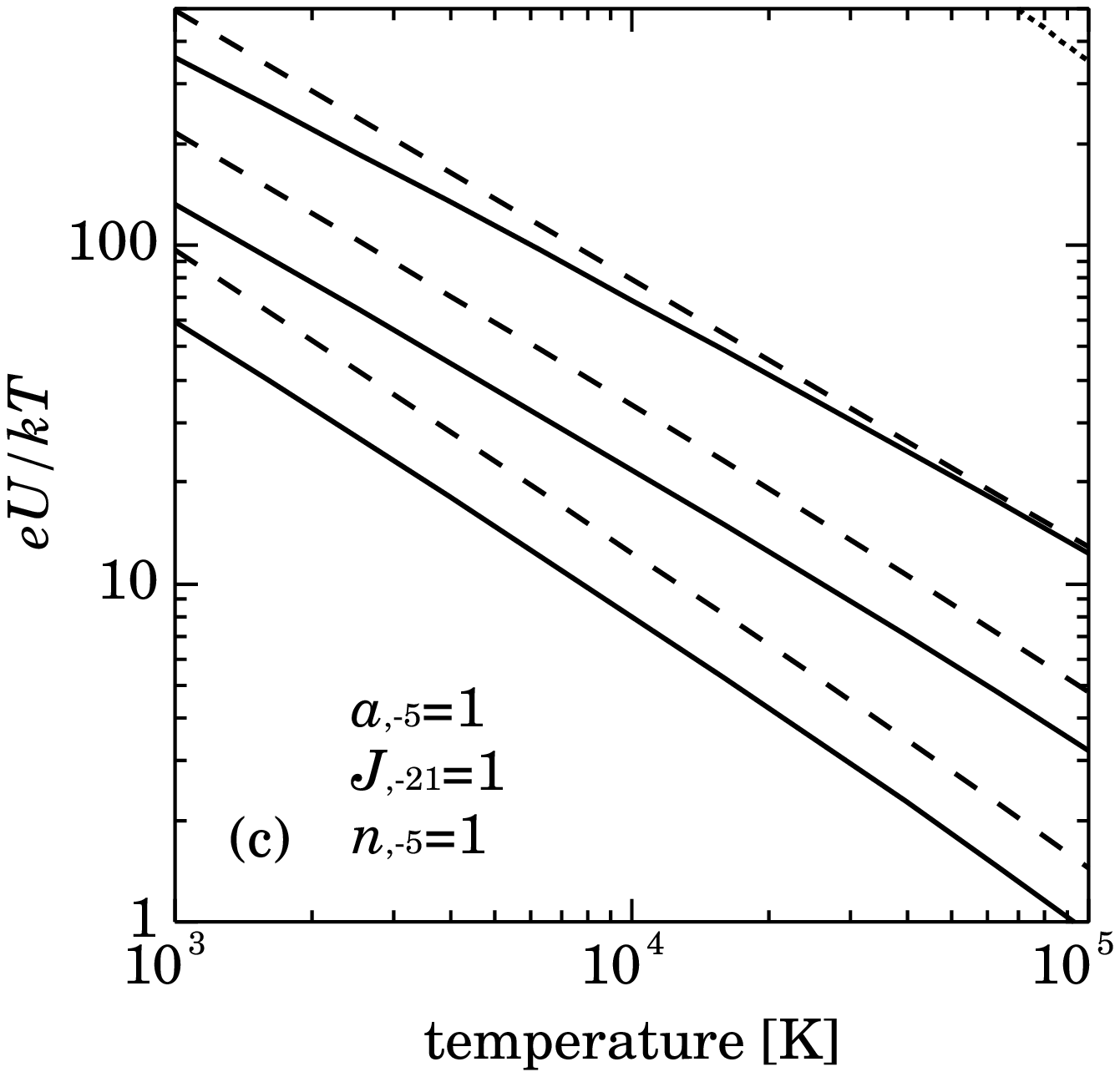}
 \includegraphics[height=8.0cm,clip] {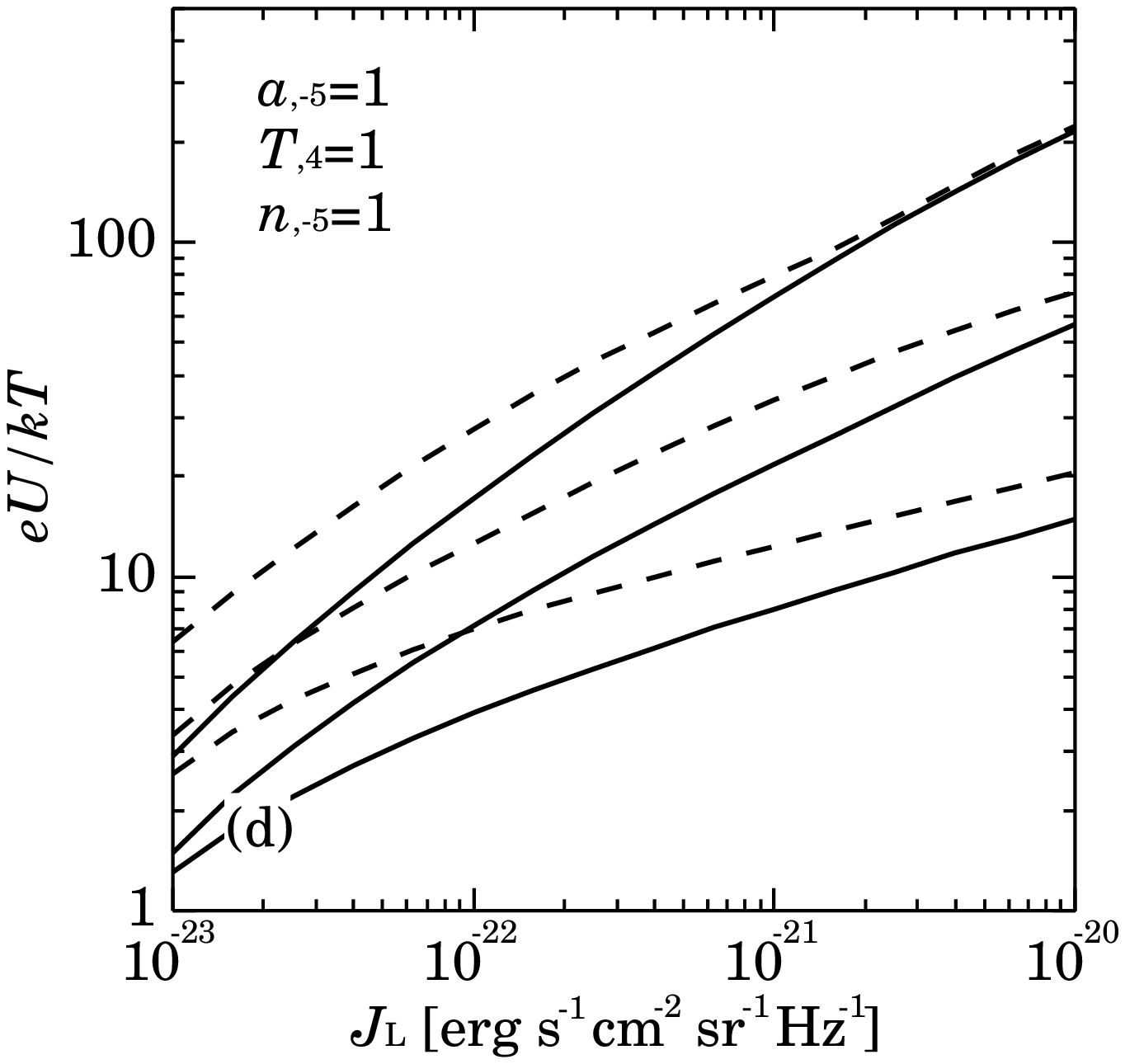}
 \caption{Normalized equilibrium potential of grains as a function of
 (a) grain size, (b) gas number density, (c) gas temperature, and (d)
 mean intensity at the Lyman limit. The solid and dashed curves indicate
 silicate and graphite cases, respectively. For the incident radiation,
 we consider three cases of the power-law index,  
 $J_\nu \propto \nu^{-\alpha}$; $\alpha=1$, 2, and 5 from top to bottom
 curves in each panel. Assumed values for parameters are indicated 
 in each panel; $a_{,-5}=a/0.1\micron$, 
 $n_{,-5}=n/10^{-5}\,{\rm cm}^{-3}$, $T_{,4}=T/10^4\,{\rm K}$, 
 and $J_{,-21}=J_{\nu_{\rm L}}/10^{-21}\,{\rm
 erg\,s^{-1}\,cm^{-2}\,sr^{-1}\,Hz^{-1}}$, 
 where $\nu_{\rm L}$ is the Lyman limit frequency. 
 The dotted lines in panels (a) and (c) indicate an upper limit of the
 grain potential \citep{dra02}.}
\end{figure}

\begin{figure}
 \includegraphics[height=8.0cm,clip] {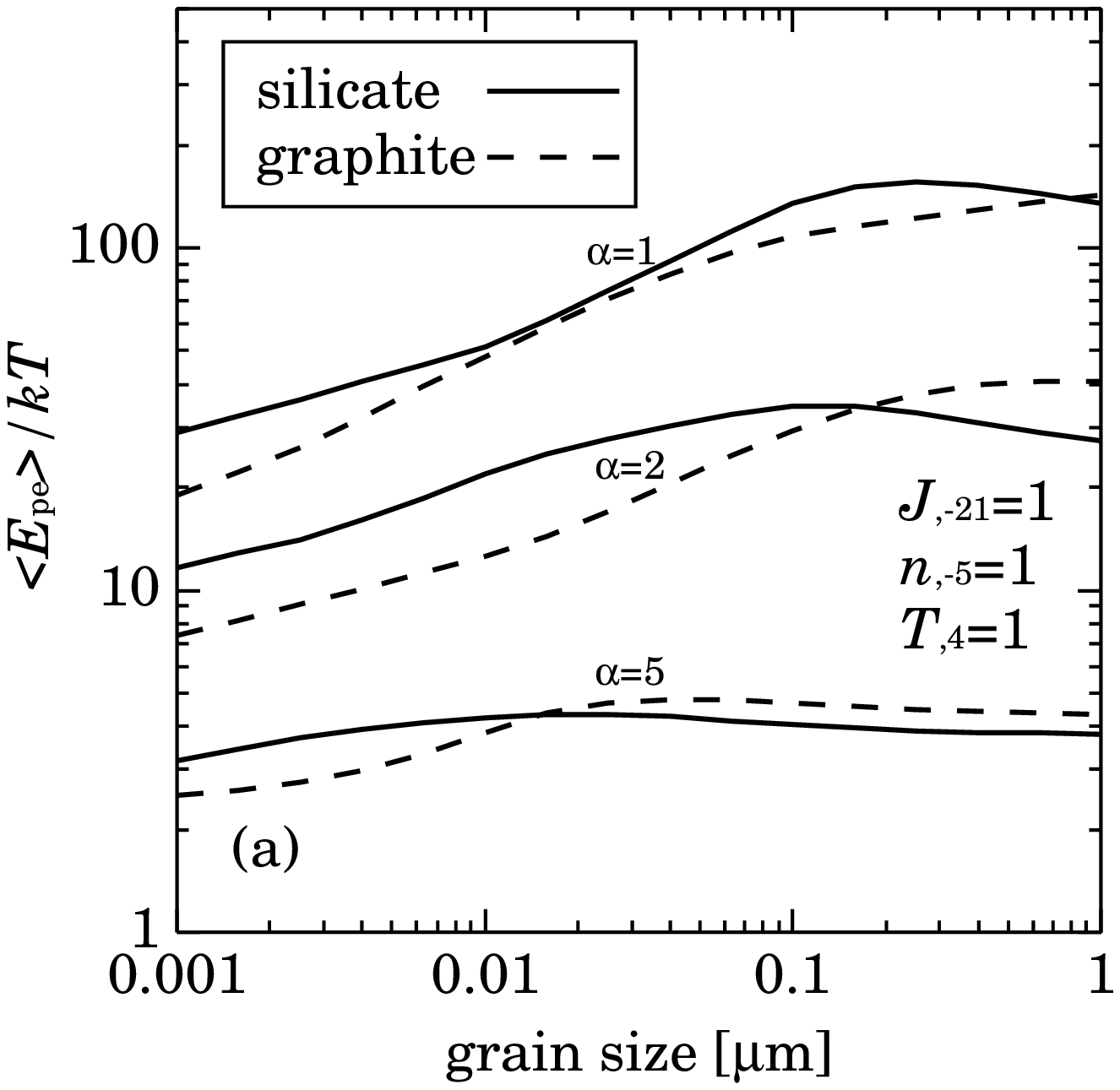}
 \includegraphics[height=8.0cm,clip] {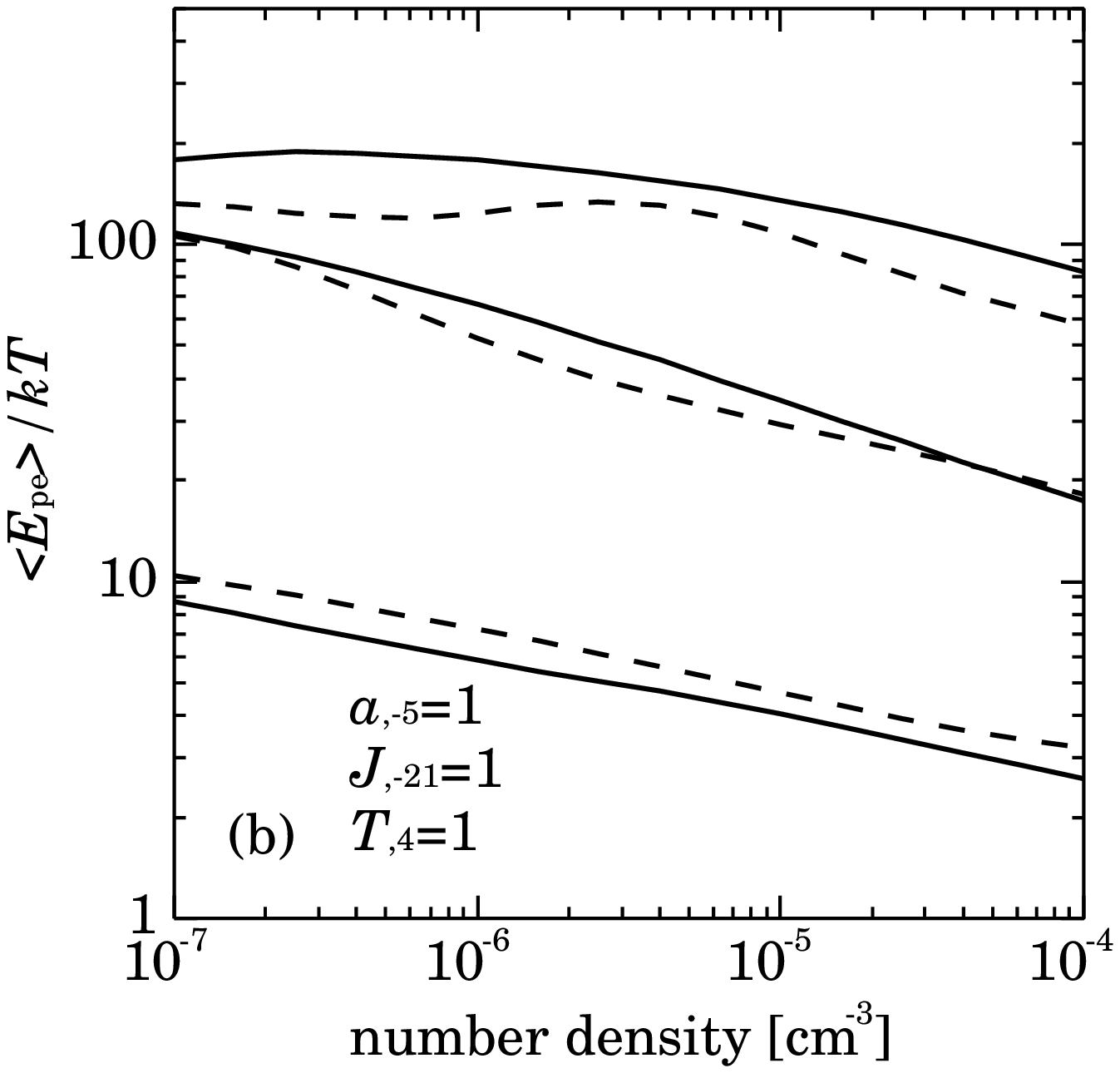}
 \includegraphics[height=8.0cm,clip] {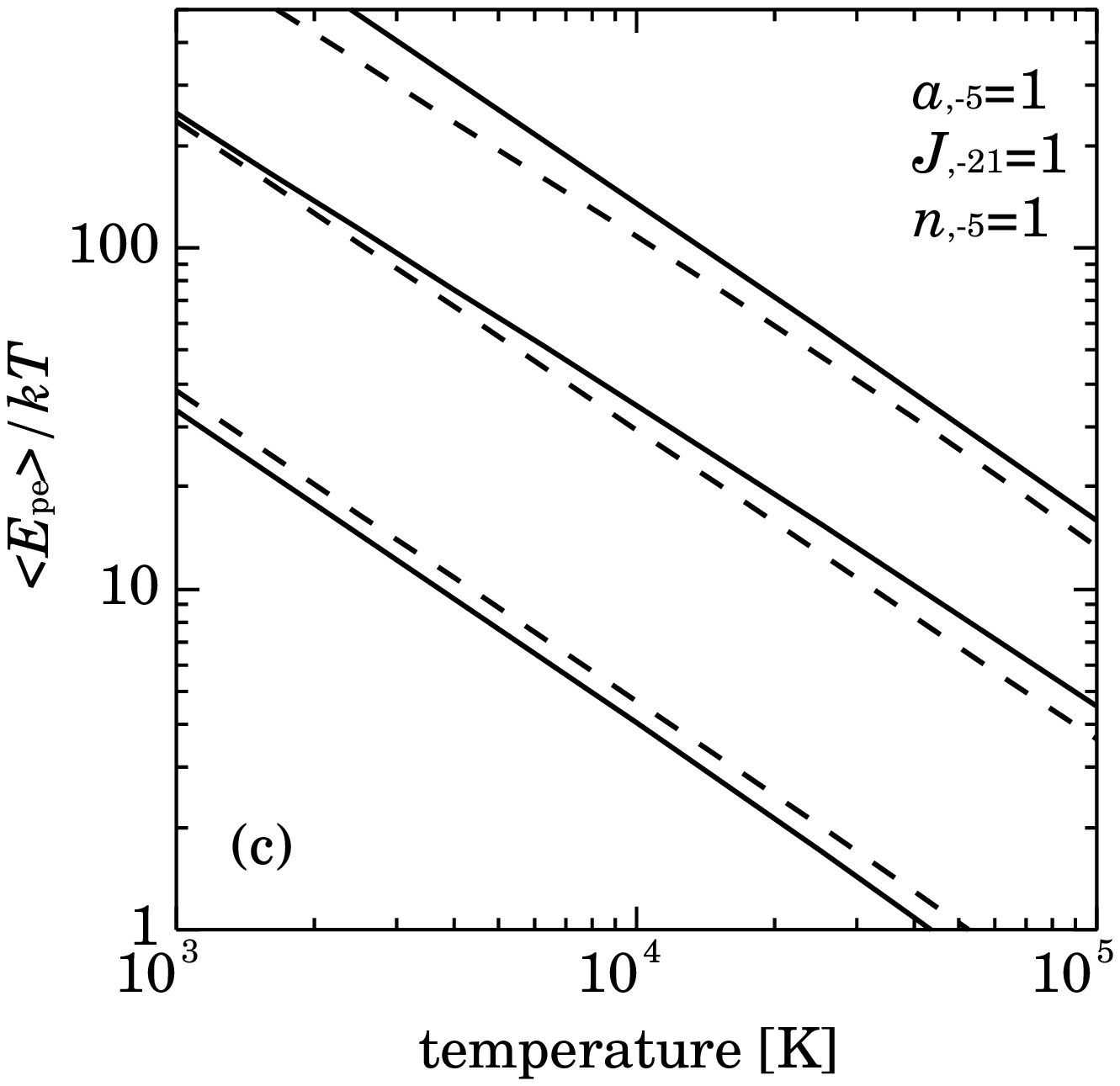}
 \includegraphics[height=8.0cm,clip] {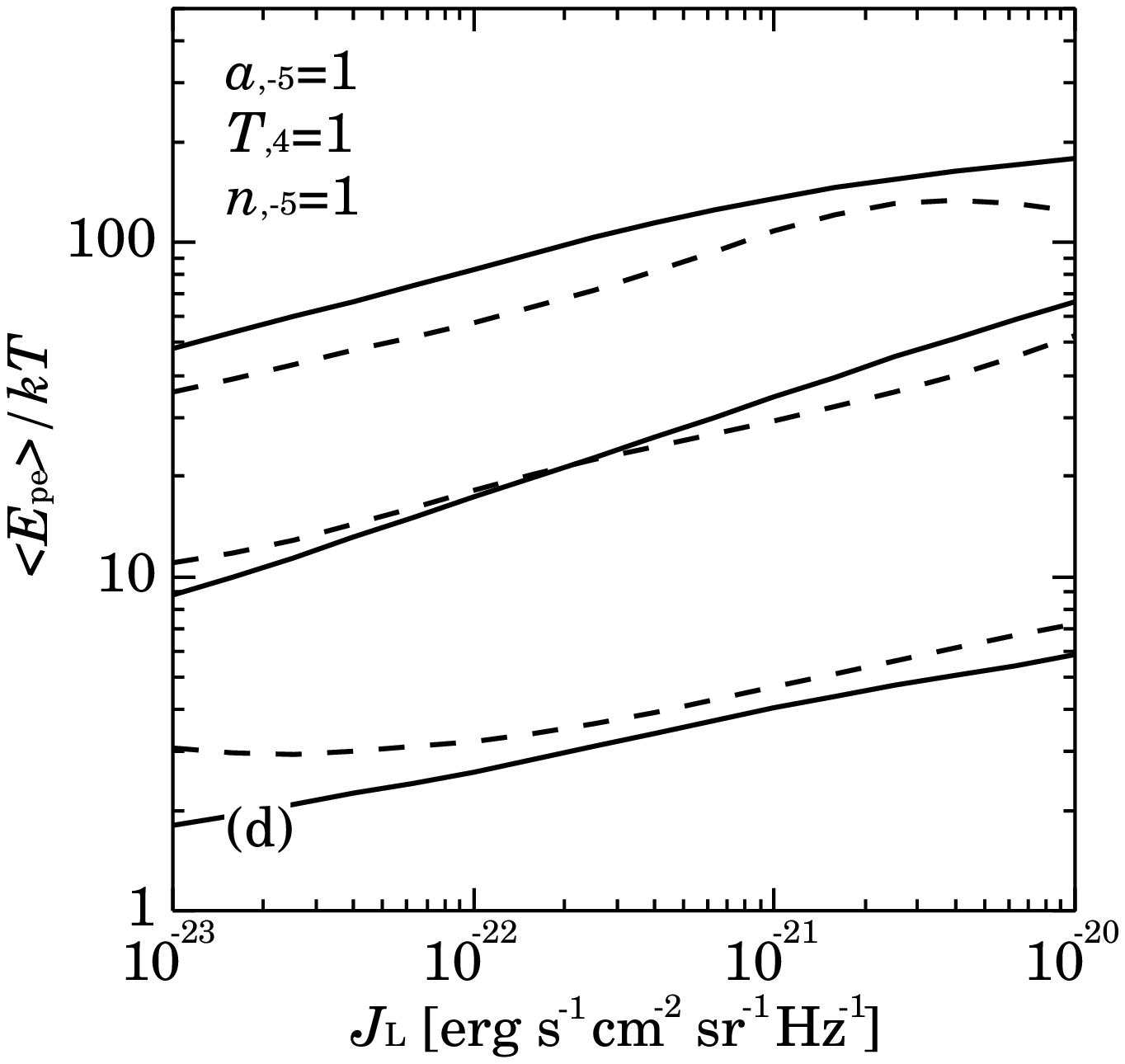}
 \caption{Mean energy of dust photoelectron normalized by gas kinetic
 energy as a function of (a) grain size, (b) gas number density, (c) gas
 temperature, and (d) mean intensity at the Lyman limit. Notations are
 the same as figure A1.}
\end{figure}

\begin{figure}
 \includegraphics[height=8.0cm,clip] {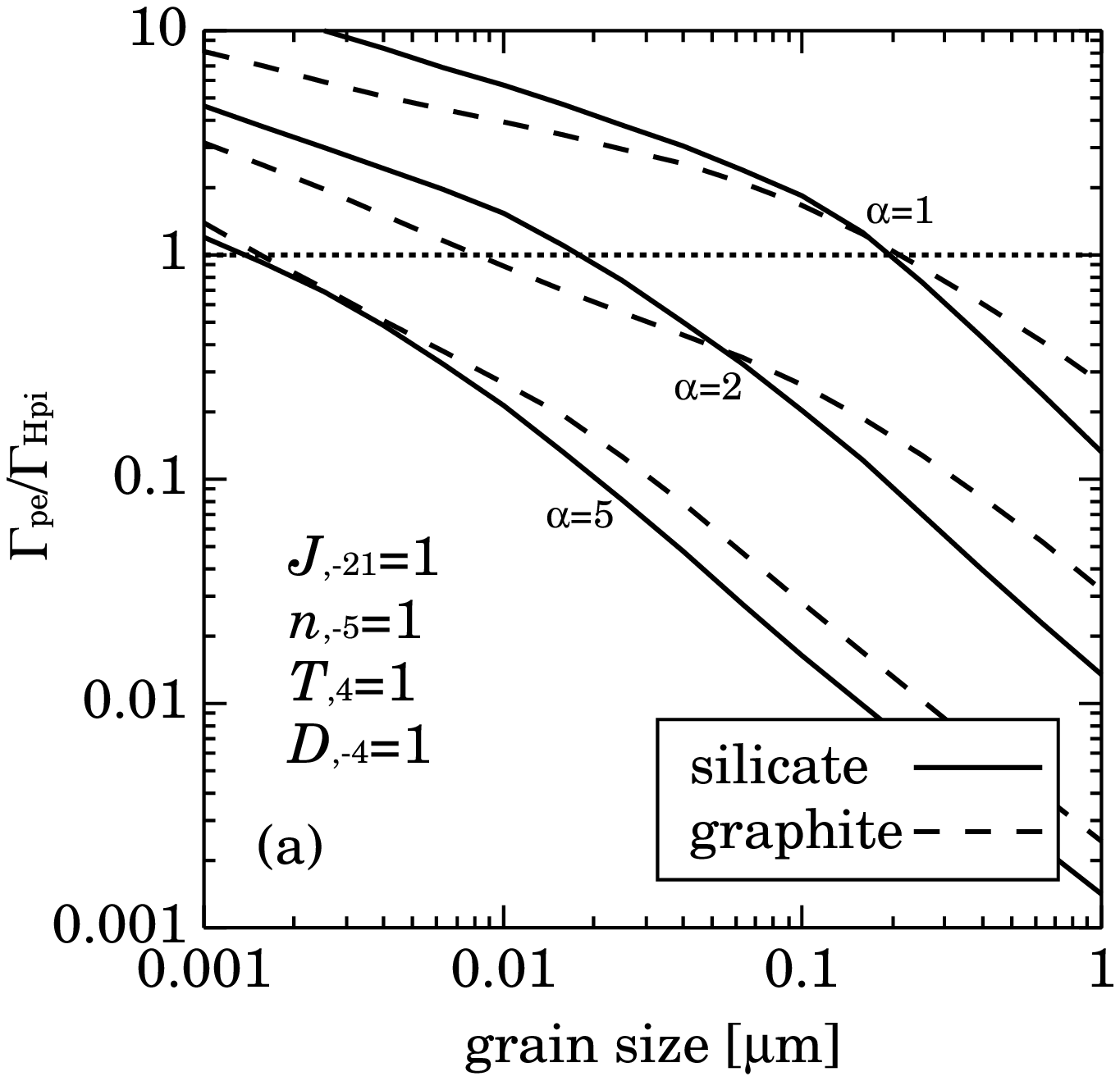}
 \includegraphics[height=8.0cm,clip] {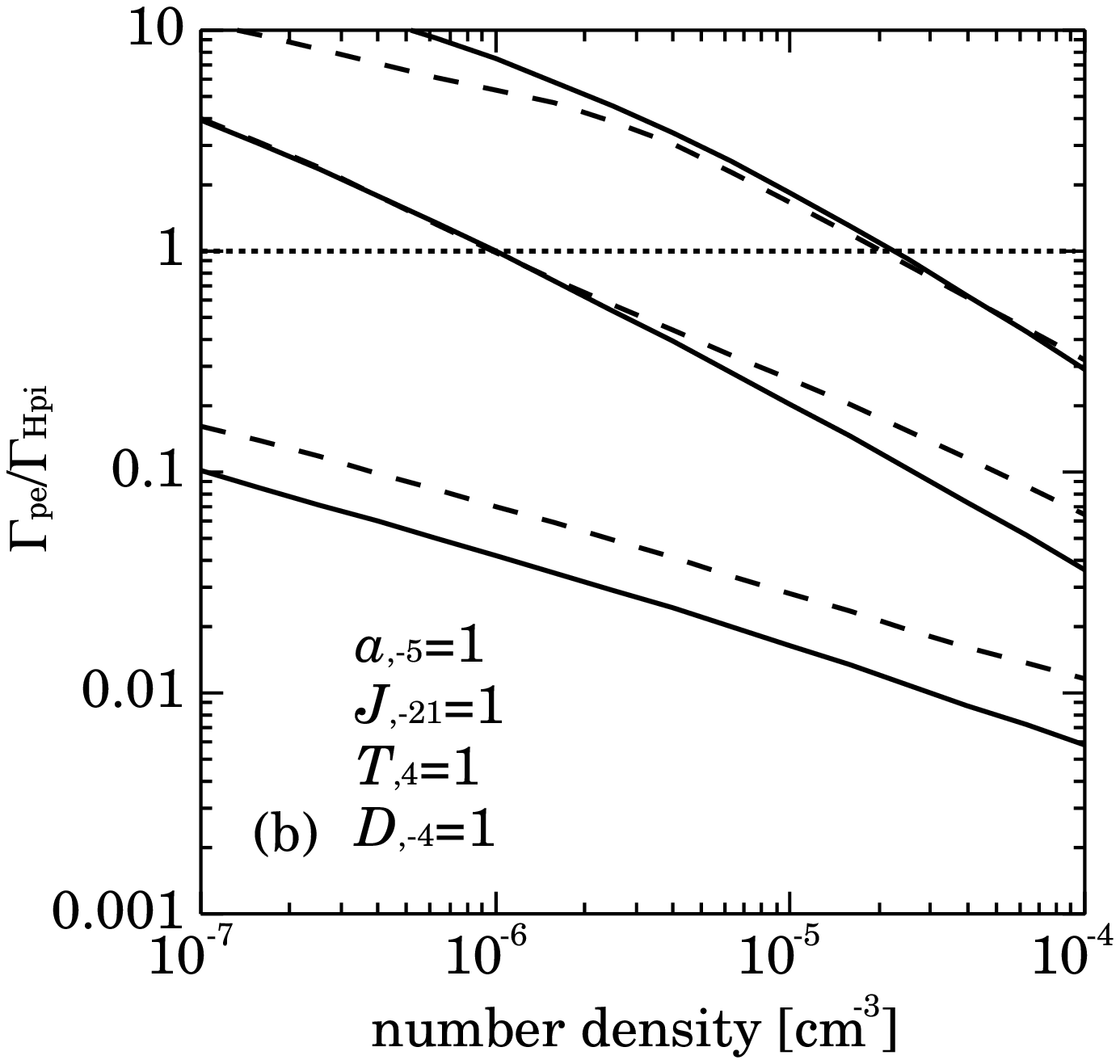}
 \includegraphics[height=8.0cm,clip] {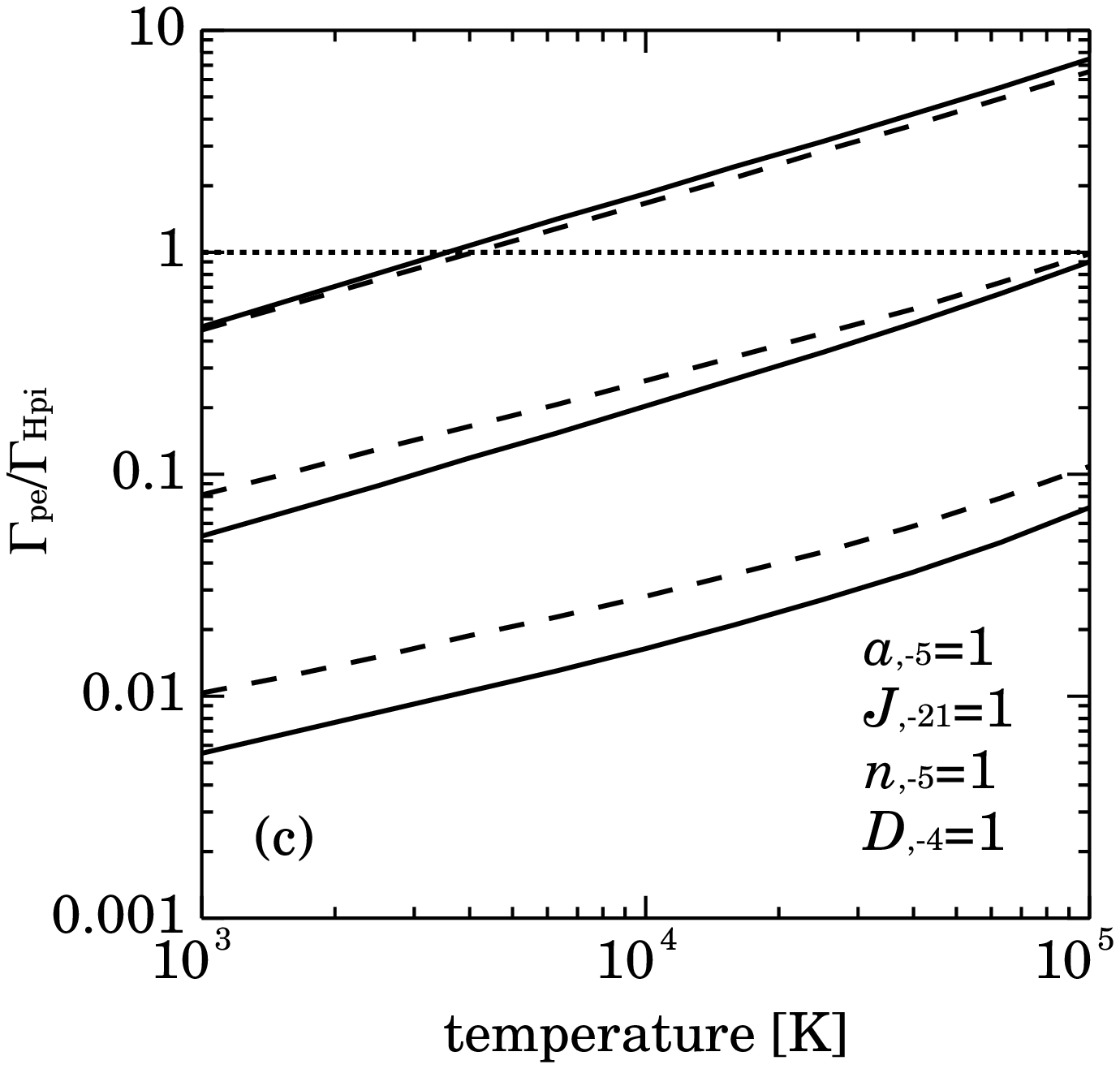}
 \includegraphics[height=8.0cm,clip] {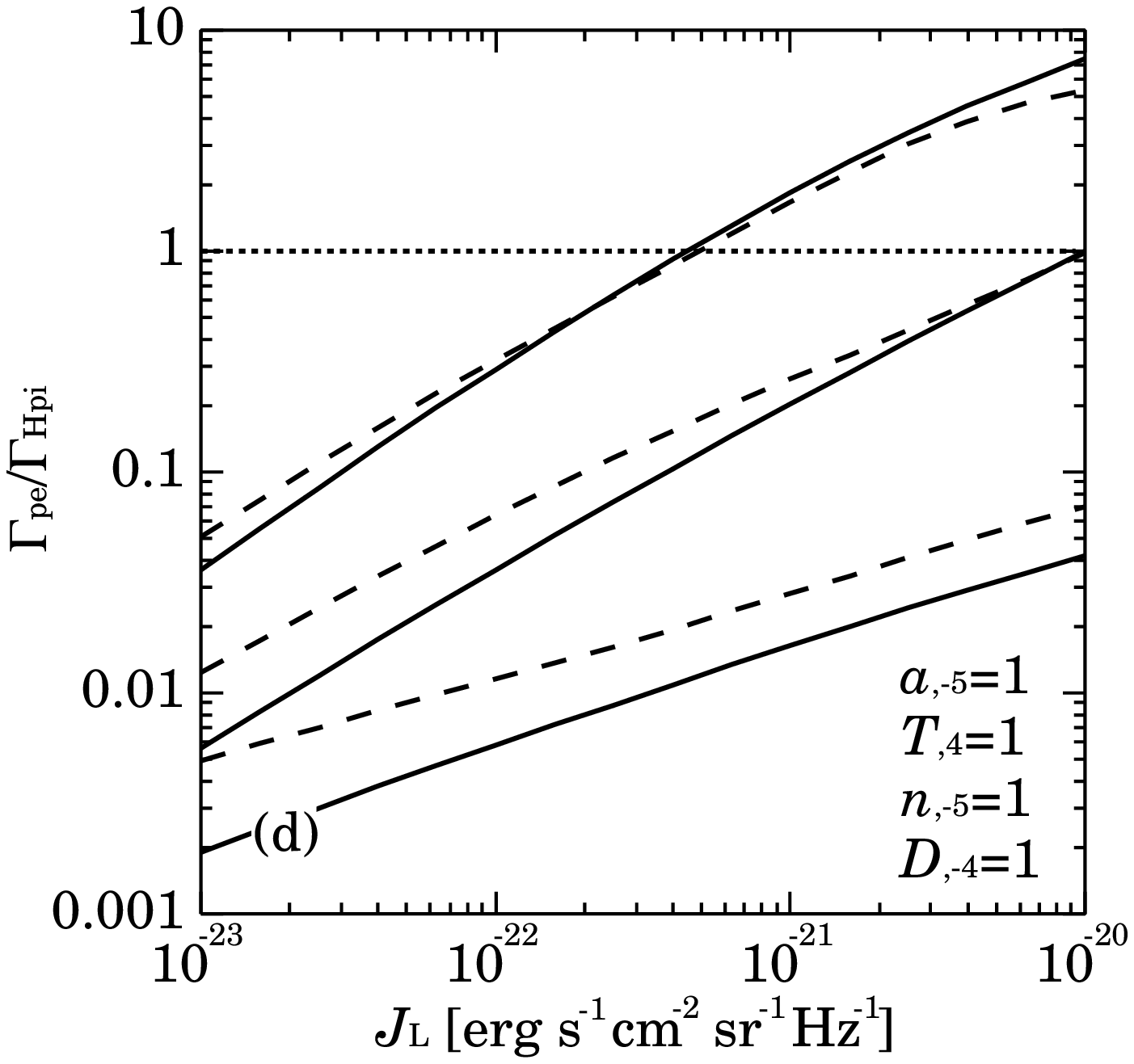}
 \caption{Ratio of dust photoelectric heating rate to hydrogen
 photoionization heating rate as a function of (a) grain size, (b) gas
 number density, (c) gas temperature, and (d) mean intensity at the
 Lyman limit. Notations are the same as figure A1. Additional parameter,
 $\cal D$ is the dust-to-gas mass ratio; ${\cal D}_{,-4}={\cal D}/10^{-4}$.}
\end{figure}

\subsection[]{Analytic investigation}

To understand the behavior of grain charge and other quantities found in
figures A1--A3 analytically, we will adopt some further approximations
in this subsection which do not appear in the method described in
\cite{ino03}.  However, the method by \cite{ino03} is used in the
calculation for making figures A1--A4 and in section 4.
We would like to ask the reader to be careful in this point. 
 
Let us express $eU/k_{\rm B}T$ as $x$. If we neglect the charging rate
by grain--ion collision, the charge equilibrium is expressed as 
\begin{equation}
 s_{\rm e} n_{\rm e} \langle v_{\rm e} \rangle (1+x)
  = \int_0^{h\nu_{\rm max}} Q_\nu Y_\nu \frac{4\pi J_\nu}{h\nu} d\nu\,,
\end{equation}
where $s_{\rm e}$ is the sticking coefficient for an electron collision,
$n_{\rm e}$ is the electron number density, $\langle v_{\rm e} \rangle$
is the mean kinetic velocity of electron, $Q_\nu$ is the absorption
efficiency factor of grains, $Y_\nu$ is the photoelectric yield, and
$J_\nu$ is the mean intensity of the incident radiation.  The left-hand
side of the above equation is the electron capture rate per unit area,
and the right-hand side is the photoelectric ionization rate per unit
area. Although the integral should be summed up to infinity in general,
we set the upper limit of the incident photon energy, 
$h\nu_{\rm max} =1.24$ keV. We also adopt $s_{\rm e}=1$.

In order to investigate analytically, 
we adopt a simple functional form for the photoelectric yield by
\cite{dra78}; 
\begin{equation}
 Y_\nu = Y_\infty \left(1-\frac{I_{\rm p}}{h\nu} \right)
\end{equation}
for $h\nu > I_{\rm p}$ and $Y_\nu = 0$ for otherwise, 
where $I_{\rm p}$ is the ionization potential;  
$I_{\rm p} \simeq W + xk_{\rm B}T$ with $W$ being the work function.
We note that a more realistic function of the photoelectric yield by 
\cite{wei01b} is used in \cite{ino03} and in section 4.  
Moreover, we adopt an approximation form of the absorption efficiency
factor as  
$Q_\nu \approx Q_{\rm L}(\nu/\nu_{\rm L})^{-\beta}$, where $\nu_{\rm L}$
is the Lyman limit frequency of hydrogen. 
For a grain larger than $\sim 0.1$ \micron, $\beta\simeq 0$, and for a
smaller grain, $\beta\simeq 1$--2 against ultraviolet photons. 
A power-law spectrum for the radiation, 
$J_\nu=J_{\rm L}(\nu/\nu_{\rm L})^{-\alpha}$ is also assumed.

If we define a function as 
$f(h\nu)=QYJ/h\nu$, $\partial f/\partial h\nu=0$ only when  
$h\nu = h\nu_* = I_{\rm p}(\alpha+\beta+2)/(\alpha+\beta+1)$.
Because the function $f$ has the single peak at $h\nu_*$, the integral in
equation (A1) is an order of $(4\pi/h)Q(h\nu_*)Y(h\nu_*)J(h\nu_*)$.
If $x\gg 1$, then, equation (A1) is reduced to 
\begin{equation}
 x^{1+\alpha+\beta} 
  \sim \frac{4\pi Q_{\rm L}Y_\infty J_{\rm L}}
  {s_{\rm e}n_{\rm e}h(\alpha+\beta+2)}
  \left(\frac{\alpha+\beta+1}{\alpha+\beta+2}\right)^{\alpha+\beta} 
  \left(\frac{h\nu_{\rm L}}{k_{\rm B}T}\right)^{\alpha+\beta} 
  \left(\frac{\pi m_{\rm e}}{8k_{\rm B}T}\right)^{1/2} \,,
\end{equation}
where we approximated $I_{\rm p} \approx xk_{\rm B}T$ because 
$W/k_{\rm B}T \sim 1$ for $T\sim 10^4$ K, and substituted 
$\langle v_{\rm e} \rangle=(8k_{\rm B}T/\pi m_{\rm e})^{1/2}$. 
If $n_{\rm e}\simeq n$ with $n$ being the gas number density, 
we find 
\begin{equation}
 \frac{eU}{k_{\rm B}T} \propto 
  \left(\frac{J_{\rm L}Q_{\rm L}(a)}{n}\right)^{1/(1+\alpha+\beta)}
  T^{-(1/2+\alpha+\beta)/(1+\alpha+\beta)}\,.
\end{equation}
Indeed, such dependences are found in figure
A1. Moreover, $Q_{\rm L} \sim (a/0.1\micron)$ for $a \la 0.1$ \micron\
and $Q_{\rm L} \sim 1$ otherwise. Thus, $eU/k_{\rm B}T$ shows nearly
no dependence of grain size for a large size.

The mean photoelectron energy is defined as 
\begin{equation}
 \langle E_{\rm pe} \rangle 
  = \frac{\int E_{\rm pe} Q_\nu Y_\nu 4\pi J_\nu/h\nu d\nu}
  {\int Q_\nu Y_\nu 4\pi J_\nu/h\nu d\nu}\,,
\end{equation}
where $E_{\rm pe}$ is the energy of the photoelectron. We can express 
$E_{\rm pe} \approx \eta (h\nu-I_{\rm p})$, where
$\eta$ is a numerical factor less than unity because a part of energy of
the incident photon is converted into the phonon energy 
of the grain \citep{wei01b}. If we adopt a parabolic function for the
energy distribution function of the photoelectron as \cite{wei01b}, the
numerical factor $\eta=1/3$--1/2 depending on the energy of the incident
photon. We adopt $\eta=1/2$ below.

If we define a function as $g(h\nu)=E_{\rm pe}QYJ/h\nu$ and assume 
the functional forms adopted above for $Q$, $Y$, and $J$,  
$\partial g/\partial h\nu$ is zero  only when  
$h\nu = h\nu_{**} = I_{\rm p}(\alpha+\beta+2)/(\alpha+\beta)$. 
Thus, the integral in the numerator of equation (A5) is an order of  
$E_{\rm pe}(h\nu_{**})Q(h\nu_{**})Y(h\nu_{**})J(h\nu_{**})$.
If $x \gg 1$, we obtain 
\begin{equation}
 \frac{\langle E_{\rm pe} \rangle}{xk_{\rm B}T} 
  \sim \frac{2}{\alpha+\beta}
  \left(\frac{\alpha+\beta}{\alpha+\beta+1}\right)^{\alpha+\beta}\,.
\end{equation}
Therefore, $\langle E_{\rm pe} \rangle/k_{\rm B}T$ has a similar
parameter dependence to $eU/k_{\rm B}T$, which is observed in figure A2.

The saturation of $\langle E_{\rm pe} \rangle/k_{\rm B}T$ is seen in the
case of $\alpha=1$ for a lower density in figure A2 (b).  This may be
due to the effect of the maximum photon energy assumed in the
calculation.  We do not consider the incident photon energy higher than
1.2 keV.  For the case, $I_{\rm p}$ reaches several hundreds eV, so that
the peak energy $h\nu_{**}$ becomes nearly the maximum energy. Thus, the
above estimate may be an overestimate for such a case.

The total photoelectric heating by dust is given by $\Gamma_{\rm
pe}=n_{\rm d}\gamma$, where $n_{\rm d}$ is the grain number density and
$\gamma$ is the heating rate per a grain. 
While $n_{\rm d}\propto n a^{-3}$ for a certain dust-to-gas ratio, 
$\gamma$ is proportional to $\langle E_{\rm pe} \rangle a^2 n T^{1/2} x$ 
because the photoelectric ionization rate per a grain
balances with the electron capture rate per a grain, where $a^2$
dependence comes from the geometrical cross section of grains. We have
assumed $x \gg 1$ again.  On the other hand, the photoionization heating
rate, $\Gamma_{\rm Hpi}$, is proportional to $n^2 T^{-0.7}$ in the
photoionization equilibrium, where the temperature dependence comes from
the recombination coefficient. Therefore, we find
\begin{equation}
 \frac{\Gamma_{\rm pe}}{\Gamma_{\rm Hpi}}
  \propto a^{-1} x^2 T^{2.2}\,,
\end{equation}
where we have used the relation 
$\langle E_{\rm pe} \rangle \propto xT$ (equation [A6]).
Remembering $J$, $n$, and $T$ dependences in $x$ described in equation
(A4), we can understand $J$, $n$, and $T$ dependences shown 
in figures A3 (b), (c), and (d).  
Because $Q_{\rm L} \propto a$ for $a \la 0.1$
\micron\ and $Q_{\rm L} \sim 1$ for otherwise, we see a double power-law
dependence of $a$ in panel (a) of figure A3.

\subsection[]{Effect of the spectral break}

As shown by \cite{haa96}, the real spectrum may show a break at the He
II Lyman limit (54.4 eV), whereas we have assumed a spectrum without
break. Here this point is discussed. Assuming the power-law spectral
index ($\alpha$) is fixed all over the spectral range for simplicity,  
we multiply the intensity of the background radiation above the
He II Lyman limit by a factor of $f_{\rm HeII}$, which is called the
spectral break factor in this appendix. 
As extreme cases, $f_{\rm HeII}=1$ means that there is no break, and
$f_{\rm HeII}=0$ means that there is no photon above the He II Lyman
limit.  In all calculations, except for figure A4, 
$f_{\rm HeII}=1$ has been assumed.

Figure A4 shows the effect of $f_{\rm HeII}$ on the normalized
mean photoelectron energy from dust grains 
($\langle E_{\rm pe} \rangle/k_{\rm B}T$), which indicates the heating
efficiency per a grain. Only the silicate case is shown, but the
graphite case is very similar. For $\alpha=1$ case (dotted curves),
we observe the photoelectron energy $\langle E_{\rm pe} \rangle/k_{\rm
B}T$ decreases with decreasing $f_{\rm HeII}$. This is because the
number of high energy photons decreases if the spectral break is larger 
(i.e. smaller $f_{\rm HeII}$). According to figure 5 in \cite{haa96},
the break is significant like $f_{\rm HeII}\sim 0.1$. 
One might think that our assumption of $f_{\rm HeII}=1$ with $\alpha=1$
results in an overestimation of the dust heating, so that a larger
amount of dust may be allowed in the IGM.

However, figure 5 of \cite{haa96} also shows that the
spectral index is 0.5 rather than unity for $\la 1$ keV photons which we
are interested in. In the $\alpha=0.5$ case (solid curves of figure A4), 
we find a good quantitative agreement with the case of $f_{\rm HeII}=1$
and $\alpha=1$ (top dotted curve) if $f_{\rm HeII}=0.1$--0.3. Therefore,
our results obtained from the background spectrum of $f_{\rm HeII}=1$
and $\alpha=1$ should be quantitatively very consistent with those from
a more realistic spectrum with the He II Lyman limit break. 

\begin{figure}
 \includegraphics[height=8.0cm,clip] {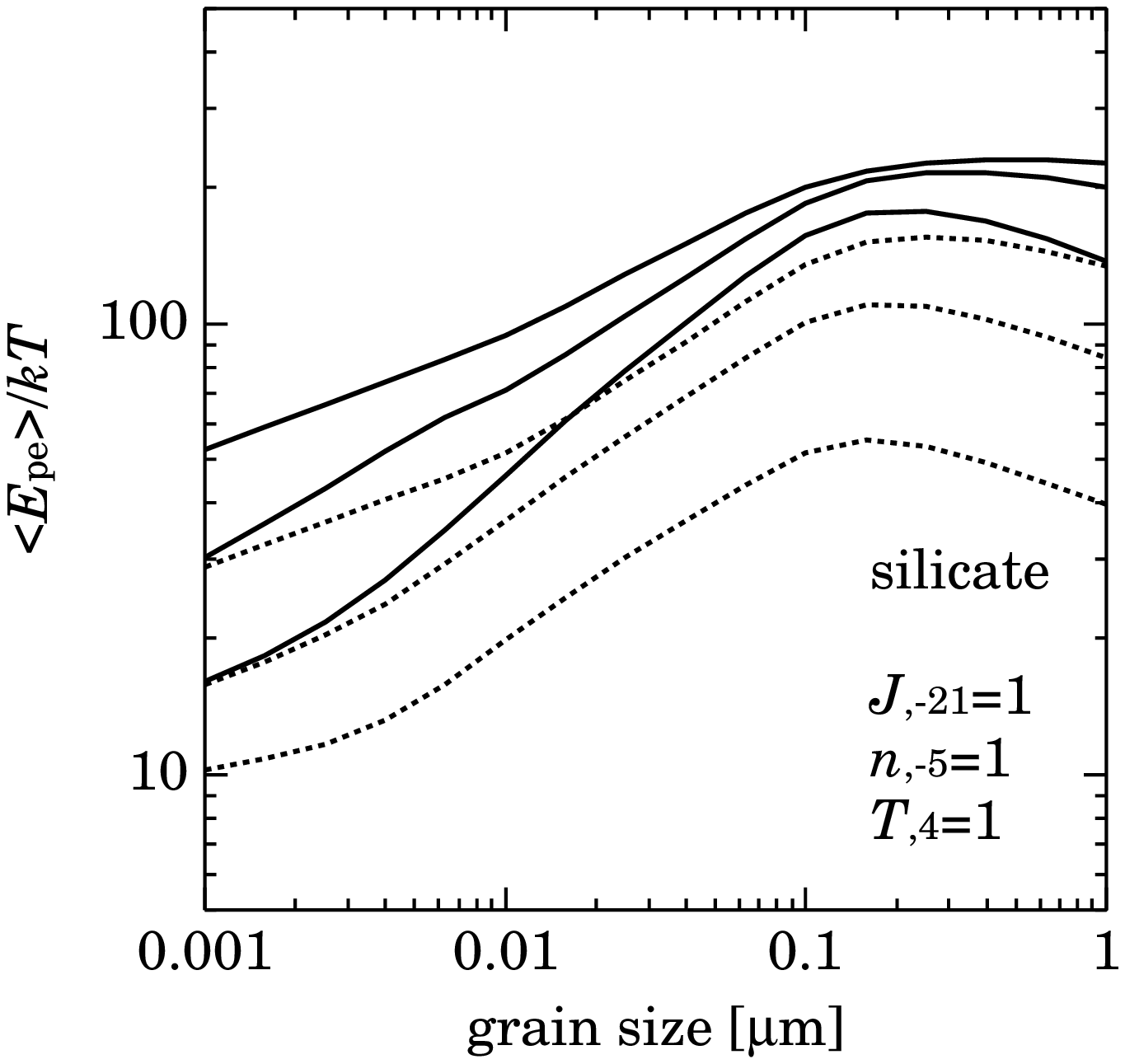}
 \caption{Effect of the spectral break at the He II Lyman limit (54.4
 eV) for silicate grains. Solid and dotted curves are the spectral index
 $\alpha=0.5$ and 1 cases, respectively. For each type of curves, we
 assume the spectral break factor, $f_{\rm HeII}=0.1$, 0.3, and 1 (no
 break) from bottom to top, respectively. The gas density 
 $n_{,-5}=n/10^{-5}\,{\rm cm}^{-3}$, the gas temperature 
 $T_{,4}=T/10^4\,{\rm K}$, and the intensity of the incident radiation 
 $J_{,-21}=J_{\nu_{\rm L}}/10^{-21}\,{\rm 
 erg\,s^{-1}\,cm^{-2}\,sr^{-1}\,Hz^{-1}}$ are assumed.}
\end{figure}

\section[]{Chemical rate equations and coefficients}

In this paper, we consider HI, HII, HeI, HeII, HeIII, and electron as
gaseous species. That is, we neglect the effect of the metal production
by stars.  The primordial helium mass fraction $Y=0.24$ is always
adopted throughout our calculation. 

Let us define a non-dimensional number abundance of each gaseous
species as   
\begin{equation}
 X_i \equiv \frac{n_i}{n_{\rm b}}\,,
\end{equation}
where $n_i$ is the number density of $i$-th species, and $n_{\rm b}$ is
the baryon number density, which is given by $n_{\rm b}(z)=\Omega_{\rm
b}\rho_{\rm c,0}(1+z)^3/\mu$ with $\rho_{\rm c,0}$ being the local
critical density, and $\mu$ being the mean mass of baryon particles.
Baryon is assumed to be only hydrogen and helium.  Using the helium mass
fraction $Y$, hence, we obtain
\begin{equation}
 X_{\rm HI}+X_{\rm HII}=\frac{4(1-Y)}{4-3Y}\,,
\end{equation}
\begin{equation}
 X_{\rm HeI}+X_{\rm HeII}+X_{\rm HeIII}=\frac{Y}{4-3Y}\,,
\end{equation}
and 
\begin{equation}
 \mu=\left(\frac{4}{4-3Y}\right)m_{\rm p}\,,
\end{equation}
where $m_{\rm p}$ is the proton mass. For electron abundance, we have 
\begin{equation}
 X_{\rm e}=X_{\rm HII}+X_{\rm HeII}+2X_{\rm HeIII}\,.
\end{equation}

Chemical rate equations for the gaseous species are 
\begin{equation}
 \frac{dX_{\rm HI}}{dt}=X_{\rm HII}X_{\rm e}n_{\rm b}\alpha_{\rm HII}
  -X_{\rm HI}(X_{\rm e}n_{\rm b}\beta_{\rm HI}+\gamma_{\rm HI})\,,
\end{equation}
\begin{equation}
 \frac{dX_{\rm HeI}}{dt}=X_{\rm HeII}X_{\rm e}n_{\rm b}\alpha_{\rm HeII}
  -X_{\rm HeI}(X_{\rm e}n_{\rm b}\beta_{\rm HeI}+\gamma_{\rm HeI})\,,
\end{equation}
and
\begin{equation}
 \frac{dX_{\rm HeIII}}{dt}=-X_{\rm HeIII}X_{\rm e}n_{\rm b}\alpha_{\rm HeIII}
  +X_{\rm HeII}(X_{\rm e}n_{\rm b}\beta_{\rm HeII}+\gamma_{\rm HeII})\,,
\end{equation}
where $\alpha_i$, $\beta_i$, and $\gamma_i$ are recombination
coefficients, collisional ionization coefficients, and photoionization
rates for $i$-th species, respectively. The adopted functions of
$\alpha_i$ and $\beta_i$ are tabulated in Table B1. 
The photoionization rate is given by
\begin{equation}
 \gamma_i=\int_{\nu_i}^{\nu_{\rm max} }\sigma_{i,\nu} 
  \frac{4\pi J_\nu}{h\nu} d\nu
  \approx \frac{4\pi}{h} \sigma_i J_{\rm L}
  \left(\frac{\nu_{\rm L}}{\nu_i}\right)^\alpha
  \left(\frac{b_i}{\alpha+s_i}
   -\frac{b_i-1}{\alpha+s_i+1}\right)\,,
\end{equation} 
where $J_\nu=J_{\rm L}(\nu/\nu_{\rm L})^{-\alpha}$ with $\nu_{\rm L}$
is the hydrogen Lyman limit frequency, and 
$\sigma_{i,\nu}$ is the ionization cross section: 
\begin{equation}
 \sigma_{i,\nu} = \sigma_i \left[
	b_i\left(\frac{\nu}{\nu_i}\right)^{-s_i}
	+(1-b_i)\left(\frac{\nu}{\nu_i}\right)^{-s_i-1}\right]\,,
\end{equation}
for $\nu\geq \nu_i$ and otherwise $\sigma_{i,\nu}=0$ \citep{ost89}.
The parameters for ionization cross sections are summarised in Table
B2. The last term in equation (B9) is valid 
when $\nu_{\rm max}\gg \nu_i$. In this paper, we set 
$h\nu_{\rm max}=1.24$ keV.

\begin{table}
 \centering
 \begin{minipage}{140mm}
  \caption{Recombination and collisional ionization coefficients in
  cm$^3$ s$^{-1}$ taken from \citet{cen92} and \citet{the98}.}
  \begin{tabular}{c}
   \hline
   Recombination \\
   $\alpha_{\rm HII}
   =6.30\times10^{-11}T^{-1/2}T_3^{-0.2}/(1+T_6^{0.7})$ \\
   $\alpha_{\rm HeII}
   =1.50\times10^{-10}T^{-0.6353}+\alpha_{\rm HeII}^{\rm D}$ \\
   $\alpha_{\rm HeIII}
   =3.36\times10^{-10}T^{-1/2}T_3^{-0.2}/(1+T_6^{0.7})$ \\
   Dielectric recombination\\
   $\alpha_{\rm HeII}^{\rm D}=1.9\times10^{-3}T^{-1.5}e^{-4.7\times10^5/T}
   (1+0.3e^{-9.4\times10^4/T})$\\
   Collisional ionization\\
   $\beta_{\rm HI}
   =5.85\times10^{-11}T^{1/2}(1+T_5^{1/2})^{-1}e^{-157809.1/T}$\\
   $\beta_{\rm HeI}
   =2.38\times10^{-11}T^{1/2}(1+T_5^{1/2})^{-1}e^{-285335.4/T}$ \\
   $\beta_{\rm HeII}
   =5.68\times10^{-12}T^{1/2}(1+T_5^{1/2})^{-1}e^{-631515/T}$ \\
   \hline
  \end{tabular}
 \end{minipage}
\end{table}

\begin{table}
 \centering
 \begin{minipage}{140mm}
  \caption{Parameters for ionization cross sections from \citet{ost89}.}
  \begin{tabular}{ccccc}
   \hline
   species & $\sigma_i$ [$10^{-18}$ cm$^2$]& $\nu_i$ [$10^{15}$ Hz]
   & $b_i$ & $s_i$ \\
   HI & 6.30 & 3.29 & 1.34 & 2.99 \\
   HeI & 7.83 & 5.94 & 1.66 & 2.05 \\
   HeII & 1.58 & 13.2 & 1.34 & 2.99 \\
   \hline
  \end{tabular}
 \end{minipage}
\end{table}

By solving equations (B6)--(B8), we obtain $X_{\rm HI}$, $X_{\rm HeI}$,
and $X_{\rm HeIII}$. Once these fractional abundances are obtained,
other abundances are found from equations (B2), (B3), and (B5).
In addition, the numbers of hydrogen and helium nuclei are constant
since the helium mass fraction $Y$ is now constant. Thus,
$dX_{\rm HII}/dt=-dX_{\rm HI}/dt$ and 
$dX_{\rm HeII}/dt=-dX_{\rm HeI}/dt-dX_{\rm HeIII}/dt$. Therefore, 
the term of $dX/dt$ in equation (6) of section 4.1 is reduced to 
$-dX_{\rm HI}/dt-dX_{\rm HeI}/dt+dX_{\rm HeIII}/dt$ because  of 
$dX/dt=\sum_i dX_i/dt=dX_{\rm e}/dt$ and equation (B5).

We now consider the atomic photoionization heating and the photoelectric
heating by dust as the heating mechanism of gas in equation (6).
The photoionization heating rate per a $i$-th species atom/ion is given
by  
\begin{equation}
 \epsilon_i = \int_{\nu_i}^{\nu_{\rm max}} (h\nu-h\nu_i)
  \sigma_{i,\nu} \frac{4\pi J_\nu}{h\nu} d\nu
  \approx 4\pi \sigma_i \nu_i J_{\rm L}
  \left(\frac{\nu_{\rm L}}{\nu_i}\right)^\alpha
  \left(\frac{b_i}{\alpha+s_i-1}
   -\frac{2b_i-1}{\alpha+s_i}
   +\frac{b_i-1}{\alpha+s_i+1}\right)\,,
\end{equation}
where parameters, $\sigma_i$, $\nu_i$, $b_i$, and $s_i$ are summarised
in Table B2, and $\alpha$ is the power-law spectral index of the
incident radiation. Again, the last term of equation (B11) is valid when
$\nu_{\rm max}\gg\nu_i$. The dust photoelectric heating is given by
calculating the equilibrium charge and the ejection rate of the
photoelectron by the manner described in \citet{ino03}.
Finally, the adopted cooling rates are summarised in Table B3. 
The metallic line cooling is not so important for our problem because the
temperature interested is less than 25,000 K and the metallicity is
1/10--1/100 of the Solar value \citep{sut93}.

\begin{table}
 \centering
 \begin{minipage}{140mm}
  \caption{Cooling coefficients in erg cm$^3$ s$^{-1}$ taken from
  \citet{cen92}, \citet{the98}, and \citet{ino03}.}
  \begin{tabular}{cc}
   \hline
   Recombination cooling & \\
   HII &
   $8.70\times10^{-27}T^{1/2}T_3^{-0.2}/(1+T_6^{0.7})X_{\rm e}X_{\rm HII}$\\
   HeII &
   $1.55\times10^{-26}T^{0.3647}X_{\rm e}X_{\rm HeII}$ \\
   HeIII &
   $3.48\times10^{-26}T^{1/2}T_3^{-0.2}/(1+T_6^{0.7})X_{\rm e}X_{\rm HeIII}$\\
   dust\footnote{$R_{\rm e}$ is the electron capture rate per a grain and
   $n_{\rm d}$ is the grain number density. These quantities are
   calculated by the method described in \citet{ino03}. Although the
   factor (3/2) is an approximation, this electron capture cooling is
   not important. A more realistic case is investigated in \cite{dra87}.}&
   $(3/2)k_{\rm B}TR_{\rm e}n_{\rm d}/n_{\rm b}^2$\\ 
   Dielectric recombination cooling & \\
   HeII & 
   $1.24\times10^{-13}T^{-1.5}e^{-4.7\times10^5/T}(1+0.3e^{-9.4\times10^4/T})X_{\rm e}X_{\rm HeII}$\\
   Collisional ionization cooling & \\
   HI & 
   $1.27\times10^{-21}T^{1/2}(1+T_5^{1/2})^{-1}e^{-157809.1/T}X_{\rm e}X_{\rm HI}$\\
   HeI & 
   $9.38\times10^{-22}T^{1/2}(1+T_5^{1/2})^{-1}e^{-285335.4/T}X_{\rm e}X_{\rm HeI}$ \\
   HeII & 
   $4.95\times10^{-22}T^{1/2}(1+T_5^{1/2})^{-1}e^{-631515/T}X_{\rm e}X_{\rm HeII}$ \\
   Collisional excitation cooling & \\
   HI & 
   $7.5\times10^{-19}(1+T_5^{1/2})^{-1}e^{-118348/T}X_{\rm e}X_{\rm HI}$\\
   HeII & 
   $5.54\times10^{-17}T^{-0.397}(1+T_5^{1/2})^{-1}e^{-47638/T}X_{\rm e}X_{\rm HeII}$\\
   Bremsstrahlung cooling & \\
   ion\footnote{$g_{\rm f}$ is the Gaunt factor and assumed to be 1.5.} & 
   $1.42\times10^{-27}g_{\rm f}T^{1/2}X_{\rm e}(X_{\rm HII}+X_{\rm HeII}+4X_{\rm HeIII})$\\
   Inverse compton cooling & \\
   CMB photon\footnote{$T_{\rm CMB}$ is the current temperature of the
   cosmic microwave background and set to be 2.7 K.} & 
   $5.406\times10^{-36}[T-T_{\rm CMB}(1+z)](1+z)^4X_{\rm e}/n_{\rm b}$\\
   \hline
  \end{tabular}
 \end{minipage}
\end{table}

\label{lastpage}
\end{document}